\documentclass[reprint,amsmath,amssymb,prxlife,superscriptaddress]{revtex4-2}
\usepackage{graphicx}% Include figure files
\usepackage{dcolumn}% Align table columns on decimal point
\usepackage{bm}% bold math
\usepackage{graphics,graphicx,epsfig}
\usepackage{epsf,epstopdf,wrapfig}
\usepackage{amsmath,amssymb}
\usepackage{xcolor}
\usepackage{dsfont}

\usepackage{newfloat}
\usepackage{algorithm}
\usepackage{algpseudocode}
\usepackage{mathtools}

\begin{document}

\title{Fluctuating landscapes and heavy tails in animal behavior}
\author{Antonio Carlos Costa}\thanks{Corresponding author. Current address: Sorbonne University, Paris Brain Institute (ICM), Inserm U1127, CNRS UMR 7225, Paris, France}
\address{
Laboratoire de Physique de l’Ecole normale supérieure, ENS, Université PSL, CNRS, Sorbonne Université, Université de Paris, F-75005 Paris, France
}%antonioccosta.phys@gmail.com}

\author{Gautam Sridhar}
\address{
Sorbonne University, Paris Brain Institute (ICM), Inserm U1127, CNRS UMR 7225, Paris, France
}
\author{Claire Wyart}
\address{
Sorbonne University, Paris Brain Institute (ICM), Inserm U1127, CNRS UMR 7225, Paris, France
}
\author{Massimo Vergassola}
\address{
Laboratoire de Physique de l’Ecole normale supérieure, ENS, Université PSL, CNRS, Sorbonne Université, Université de Paris, F-75005 Paris, France
}

\date{\today}

\begin{abstract}
    Animal behavior is shaped by a myriad of mechanisms acting on a wide range of scales, which hampers quantitative reasoning and the identification of general principles.  Here, we combine data analysis and theory to investigate the relationship between  behavioral plasticity and heavy-tailed statistics often observed in animal behavior. Specifically, we first leverage high-resolution recordings of \emph{C. elegans} locomotion to show that stochastic transitions among long-lived behaviors exhibit heavy-tailed first passage time distributions and correlation functions. Such heavy tails can be explained by slow adaptation of behavior over time. This particular result motivates our second step of introducing a general model where we separate fast dynamics on a quasi-stationary multi-well potential, from non-ergodic, slowly varying modes. We then show that heavy tails generically emerge in such a model, and we provide a theoretical derivation of the resulting functional form, which can become a power law with exponents that depend on the strength of the fluctuations. Finally, we provide direct support for the generality of our findings by testing them in a {\it C. elegans} mutant where adaptation is suppressed and heavy tails thus disappear, and recordings of larval zebrafish swimming behavior where heavy tails are again prevalent.
\end{abstract}

\maketitle

\section{Introduction} 

Animals continuously sense, process sensory information, and respond appropriately to ensure survival. High-dimensionality and multiple timescales of these far-from-equilibrium systems challenge quantitative understanding. Yet, recent advances in machine vision technologies (e.g., \cite{Pereira2020,Mathis2020,Hebert2021}) make it possible to record an animal's pose in unconstrained environments with unprecedented resolution. Such data now span several orders of magnitude \cite{Berman2018}, motivating modeling approaches that can bridge from sub-second movements to hours-long strategies.

Despite these technical advances, a complete microscopic description is not available and, most likely, out of reach. Indeed, that would require the current posture of the animal together with its physiological, sensory, and motor state\,: the uncountable number of molecules involved makes it unrealistic to track them all. Progress relies on the educated hope that so many details are not needed, as selected statistical physics examples illustrate \cite{Sethna2006}. To wit, an effective equation for the slowly varying concentration field is sufficient to capture how odor molecules diffuse in the air \cite{crank_mathematics_1979}. Much of statistical mechanics relies on the identification of such slowly varying macroscopic modes, which, through a time-scale separation, depend only statistically on microscopic details. Identifying macroscopic modes may not be a simple task, though, and it requires intuition often not immediate for far-from-equilibrium systems as encountered in biology. Here, we leverage the notion of slowly varying collective variables to motivate our introduction of reduced-order models directly from imaging data of behaving animals.

\begin{figure*}
    \centering
    \includegraphics{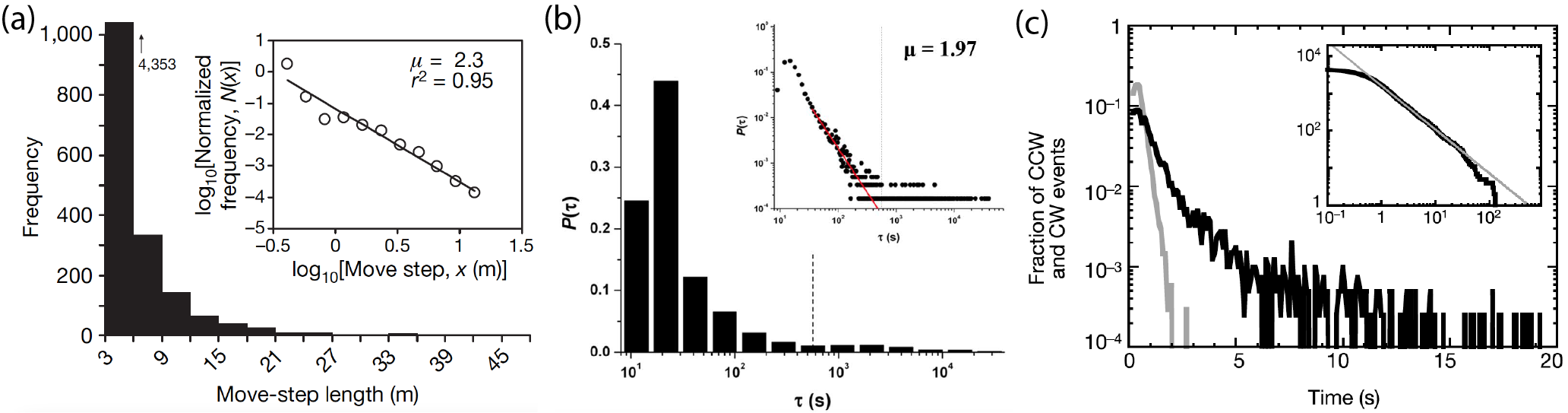}
    \caption{{\bf Power law distributions observed across species.}
    (a) Distribution of step lengths $\ell$ for an individual basking shark ({\it Cetorhinus maximus}) (adapted from Ref.~\cite{Sims2008}). The inset shows the probability density $f(\ell)$ in a $\log$-$\log$ scale, and the power law fit $\ell^{-\mu}$ with $\mu = 2.3$. Assuming a constant speed, the time spent in a step would also be distributed as $f(t) \approx t^{-\mu}$.
    (b) Distribution of the time between decisions in a choice task by Sprague Dawley rats (adapted from Ref.~\cite{Jung2014}). The inset shows the same curve in $\log$-$\log$ scale, and the power law fit $t^{-\mu}$ with $\mu = 1.97$.
    (c) Distribution of the duration of clockwise (gray) and counterclockwise (black) rotations of a single \emph{E. coli} motor (adapted from Ref.~\cite{Korobkova2004}). The inset shows the complementary cumulative distribution function (black) with a superposed power law $t^{-1}$ (gray), corresponding to a probability density $\sim t^{-2}$.
    }
    \label{fig:1}
\end{figure*}
Our starting point is the nematode \emph{C. elegans}, a pivotal model organism \cite{Brenner1974,Bargmann2013}. On a two-dimensional agar plate, worms move by propagating dorsoventral waves throughout their bodies and controlling their frequency, wavelength, and direction to move forward, backward, or turn. Long sequences of such short-lived movements exhibit signatures of chaos \cite{Ahamed2021,Costa2023}. Despite this inherent variability, time-delay embedding \cite{Takens1981,Sugihara1990,Sauer1991,Stark1999,Stark2003} yields a high-fidelity Markov model that predicts \emph{C. elegans} foraging behavior \cite{Costa2023markovian}. The resulting simulated worms are nearly indistinguishable from real ones across a wide range of scales. This Markov model also directly recovered long-lived metastable states that correspond to transitions between relatively straight paths (``runs'') and not-so-abrupt reorientations (``pirouettes'') \cite{Pierce-Shimomura1999,Costa2023markovian} (akin to the run-and-tumbling of bacteria \cite{Fujiwara2002,Berg2004}), thus providing an effective coarse-grained description of the dynamics. 

Empirical evidence for the emergence of stereotypy in the dynamics of {\it C. elegans} reflects the timescale separation between short-term movements in a given behavioral state, and long-term transitions between states. Here, we make this evocative picture concrete by recasting it in terms of models of particles hopping among wells in a potential landscape. In the first section, we build an effective Langevin description for the inferred ``run-and-pirouette'' dynamics. Notably, we find long-range correlations and heavy-tailed distributions of residence times spent in the two states (instead of the exponential expected for independent transition events). Our observation adds to the body of evidence showing that times spent in a given behavioral state are often heavy-tailed. Indeed, power-law distributions $f(t) = t^{-2}$ (where $f(t)$ is the probability density of observing a residence time of duration $t$) are found extensively across species, see Fig.\,\ref{fig:1}. In the context of search behavior, such observations have led to the hypothesis that L\'evy flights (with an exponent $-2$) result in efficient search strategies and are then evolutionarily selected \cite{Viswanathan1999,Wosniack2015,Wosniack2017,Guinard2021,Clementi2021}.

Our goal here is to combine data analysis and theory to show that heavy-tailed distributions can emerge from a slow adaptation of behavior over time. First, we infer time-dependent model parameters from the worm data and show that the observed heavy-tailed distributions can be explained by slow adaptation. Then, we introduce a general model and analyze it theoretically to account for the ubiquitous observation of heavy-tailed statistics in animal behavior. The model features potential landscapes that slowly fluctuate in time, and we demonstrate heavy-tailed first passage times and long-range correlations. The specific point that we bring here is that we obtain an analytical expression for the exponents of the power law distributions as a function of the strength of the fluctuations.  
The scaling $t^{-2}$ mentioned above is recovered as a special limiting case. Finally, the generality of our point that behavioral plasticity may be responsible for heavy tails is strengthened by the analysis of experiments on a {\it C. elegans} mutant and on larval zebrafish that confirm our predictions.

\section{Data-driven analysis reveals heavy tails in \emph{C. elegans} behavior: the role of adaptation}

\begin{figure*}
    \centering
    \includegraphics{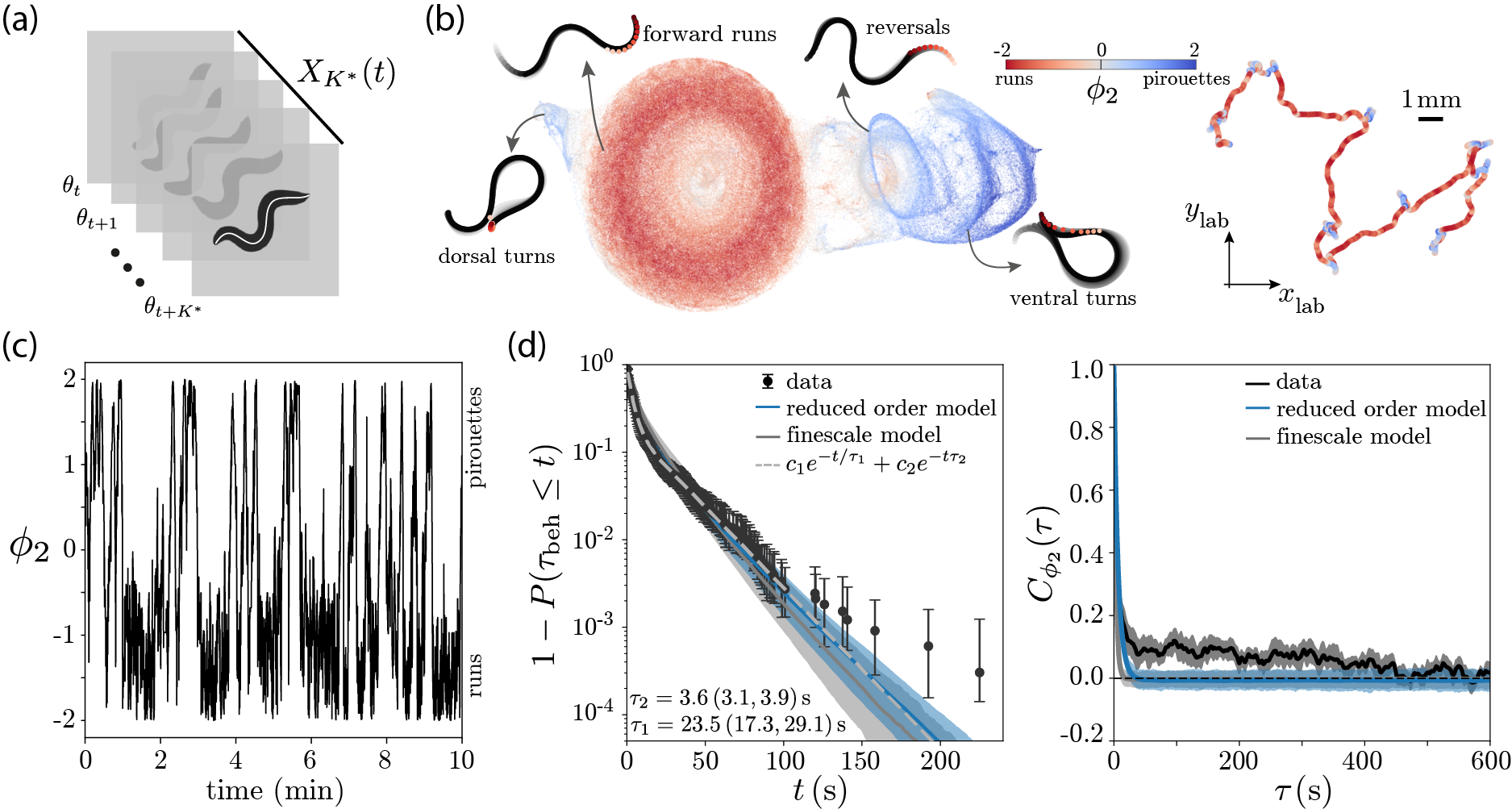}
    \caption{ {\bf A reduced-order model of \emph{C. elegans} foraging dynamics.}
    (a) From video imaging data, we measure local tangent angles along the body to obtain the body posture vector $\theta_t$ as in Ref.~\cite{Stephens2008}. A series of $K^*$ such vectors are then stacked to yield the variable $X_{K^*}(t)$ defined in the text, which captures short-term memory.
    (b) A high-fidelity Markov model of the dynamics is obtained using methods in Ref.~\cite{Costa2023markovian,Costa2023}. The first non-trivial eigenvector $\phi_2$ of the inferred Markov chain captures the long-time dynamics of the system, as discussed in the body of the paper. We represent the high-dimensional state-space $X_{K^*}$ through a 2D UMAP embedding as in \cite{Costa2023markovian} (left), and color-code each point by its projection along $\phi_2$. An example $10\,\text{min}$-long centroid trajectory color-coded by $\phi_2$ is shown on the right. The example showcases how negative/positive values of $\phi_2$ correspond to forward ``runs''/combinations of reversals, ventral and dorsal turns during ``pirouettes''.
    (c) Example time series of $\phi_2$ illustrating the stochastic hopping between ``runs'' and ``pirouettes''.
    (d) Left: Distribution function of observing a ``run'' or a ``pirouette'' with a duration longer than $\tau_\text{beh}$, $1-P(\tau_\text{beh}\leq t)$, estimated from the experimental data (black), simulations of Eq.\,\ref{eq:phi2_dot}, i.e., the dynamics projected onto $\phi_2$ (blue), and simulations of Eq.~\,\ref{eq:rho_dot}, i.e., of the full unprojected model (gray). While simulations capture the sum of exponential functions (gray dashed line) that approximates the bulk of the distribution, heavy tails observed in the data are not well captured.
    Right: Connected autocorrelation function $C_{\phi_2}(\tau)$ for the data (black) and simulations of the projected/unprojected model (blue/gray). Simulations fail again in predicting the long-range correlations exhibited by the data. Note that the projected and the full model yield similar results, illustrating the efficiency of our projection method. Error bars represent 95\% confidence intervals bootstrapped across worms.
    }
    \label{fig:2}
\end{figure*}

We leverage a previously analyzed dataset in which 12 lab-strain N2 worms are placed on an agar plate and allowed to freely explore for $T_\text{expt} = 35\,\text{minutes}$ \cite{Broekmans2016}. Our procedure is illustrated in Fig.~\ref{fig:2} and summarized hereafter, with more details deferred to Appendix A. 

From each video frame (sampled every $\delta t=1/16 s$), we extract the worm's centerline, measure tangent angles equally spaced along the body, and subtract the overall rotation of the worm to obtain the animal's posture vector $\theta_t$. Given a time $t$, the future evolution of $\theta$ does not depend on $\theta_t$ only, which reflects the effect of history and breaks Markovianity. This problem was circumvented in Refs.~\cite{Ahamed2021,Costa2023markovian} by including past postures in the description of the system. In other words, short-term memory is taken into account by expanding the state space so that it admits an approximately Markovian description. The procedure detailed in Refs.~\cite{Ahamed2021,Costa2023markovian} yields that a sequence $X_{K^*}(t) = \{\theta_{t-K^*\delta t},\ldots,\theta_{t-\delta t},\theta_t\}$ of $K^*=11$ stacked postures is sufficient to determine future statistics. The corresponding probability density $\rho_t = \rho(X_{K^*},t)$ is advanced in time by the so-called transfer operator $\mathcal{L}$\,: 

\begin{align}\label{eq:rho_dot}
    \frac{d}{dt}\rho_t &= \mathcal{L}\rho_t\,,
\end{align}
which does not need more specifics here. 

While the $\theta$ variables are continuous, it is more efficient to cluster the space of posture stacks $X_{K^*}$. Clustering yields a set of discrete states that summarize information on the dynamics. The operator $\mathcal{L}$ in Eq.~\eqref{eq:rho_dot} reduces then to a matrix with diagonal/off-diagonal entries expressing the probability to remain in the current discrete state or jump to another one. The resulting Markov chain is embodied in the transition matrix $P_{ij}(\tau)=\left(e^{{\cal L}\tau}\right)_{ij}$, which expresses the probability of transitioning between discrete states $s_i$ to $s_j$ in a time $\tau$. The procedure is detailed in \cite{Costa2023,Costa2023markovian} (see also Appendix A).

To conclude the description of our data analysis, we are left to notice that eigenvalues $\Lambda_i$ and eigenvectors $\phi_i$ of the matrix $\mathcal{L}$ and its exponential $e^{\mathcal{L}\tau}\phi_i = e^{\Lambda_i\tau}\phi_i$ provide a hierarchy of dynamical timescales \cite{Mori1965,Zwanzig1973,Rupe2022}. For a mixing system, there is a unique largest eigenvalue $\Lambda_1=0$ that corresponds to the steady-state $\phi_1$. The remaining eigenfunctions $\phi_{i>1}$, ranked by their decreasing real parts, correspond to collective variables that relax on faster and faster timescales $\sim |\mathrm{Re}(\Lambda_{i>1})|^{-1}$. 

For the measured \emph{C. elegans} foraging dynamics, the eigenspectrum of $\mathcal{L}$ reveals a slow mode $\phi_2$ that is relatively well separated from the rest of eigenmodes and coarse-grains the behavior into ``runs'' and ``pirouettes'' \cite{Costa2023markovian}, as illustrated in Fig.~\ref{fig:2}(b). Large positive/negative values of $\phi_2$ correspond to pirouettes/runs, respectively. This property allows us to define a slow reaction coordinate \cite{Froyland2014}  that captures the worm's dynamics along a ``run-and-pirouette'' axis, see Fig.\,\ref{fig:2}(b). In particular, as the worm moves, it traces an orbit in the $X_{K^*}$ space that we project onto $\phi_2$. The bottom line is that the fast dynamics of the body postures \cite{Ahamed2021} is integrated out and the projection onto $\phi_2$ highlights the effective stochastic description for the hopping between ``runs'' and ``pirouettes''. A typical time series of $\phi_2(t)$ is shown in Fig.~\ref{fig:2}(c).

\medskip
\subsection{Inferring a stationary Langevin equation for the ``run-and-pirouette'' dynamics}

To infer an explicit model for stochastic hoppings along $\phi_2(t)$, we sample the dynamics at the Markov-Einstein timescale $\tau^*$ \cite{Friedrich2011,Callaham2021}, i.e., long enough that effects of higher-order eigenmodes $\phi_i$ ($i\ge 3$) have decayed. Thus, we can obtain an effective overdamped Langevin description for $\phi_2(t)$ \footnote{We use the It\^{o} interpretation of the stochastic dynamics (see, e.g., \cite{vanKampen1981})}\,:
\begin{equation}\label{eq:phi2_dot}
    \dot{\phi_2} = F(\phi_2) + \sqrt{2 D(\phi_2)}\eta(t)\,,
\end{equation}
where we effectively have $\langle \eta(t)\eta(t')\rangle \simeq \delta(t-t')$ due to the coarse sampling every $\tau^*$. In practice, the choice $\tau^*=0.75\,\text{s}$ ensures that a stochastic model inferred from the {\it C. elegans} time series results in effectively delta-correlated fluctuations, Fig.\,S1(b). To find $F(\phi_2)$ and $D(\phi_2)$ we use a kernel-based approach \cite{Lamouroux2009} based on the Kramers-Moyal expansion \cite{Risken1989}, rather than discretized bins, to obtain a more robust estimate (see Appendix A).
 
To probe the relevance of the above model, we identify ``run'' and ``pirouette'' states by maximizing the metastability of both states (see Appendix A) \cite{Costa2023markovian}, and estimate the probability $P(\tau_\text{beh})$ of a residence time $\tau_\text{beh}$ in one of the two behaviors, see Fig.\,\ref{fig:2}(d-left) and Fig.\,S2.  Interestingly, while the exponential bulk of $P(\tau_\text{beh})$ is captured by Eq.~\ref{eq:phi2_dot}, heavier tails are not. In addition, we estimated the connected autocorrelation function 
\begin{equation}
    C_{\phi_2}(\tau) = \frac{1}{\sigma^2_{\phi_2}}\langle(\phi_2(t)-\langle\phi_2\rangle_t)(\phi_2(t+\tau)-\langle\phi_2\rangle_t)\rangle_t\,,
\end{equation}
where $\sigma^2_{\phi_2}$ is the variance of $\phi_2(t)$ and $\langle \cdot \rangle_t$ denotes temporal average. We observe again that the model captures short timescales ($\approx 10\,\text{s})$ but fails to predict long-range correlations exhibited by the data, Fig.\,\ref{fig:2}(d-right). This discrepancy is not due to the projection onto $\phi_2$ or the assumption of Langevin dynamics since simulations of the full model Eq.~\ref{eq:rho_dot} yield similar predictions, see Fig.\,\ref{fig:2}(d).

\begin{figure*}
    \centering
    \includegraphics{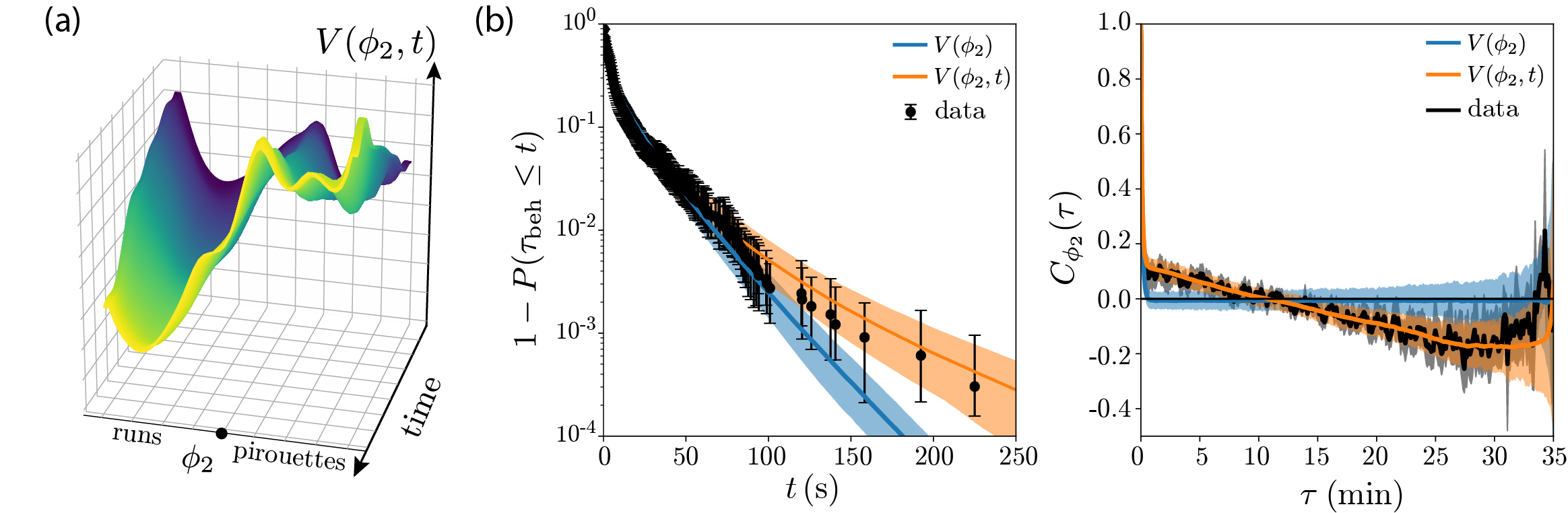}
    \caption{{\bf A time-varying potential landscape captures heavy tails in \emph{C. elegans} behavior.}
    (a) The time-dependent potential landscape for the eigenmode $\phi_2$ discussed in the text. As time goes on (blue to yellow), the barrier remains close to $\phi_2=0$ (black dot on the $\phi_2$ axis) while the ``run'' well becomes deeper over time. That witnesses adaptation towards increasingly performing ``runs''.
    (b) The probability of observing a ``run'' or a ``pirouette'' with a duration $> \tau_\text{beh}$, $1-P(\tau_\text{beh}\leq t)$ (left), and the connected autocorrelation function, $C_{\phi_2}(\tau)$ (right), as obtained from data (black), simulations of the static model Eq.\,\ref{eq:phi2_dot} (blue, same as Fig.\,\ref{fig:1}(d)), and simulations of the time-dependent model Eq.\,\ref{eq:phi2_dot_t} (orange). Note that the latter captures heavy tails and long-range correlations observed in the data. Error bars represent 95\% confidence intervals bootstrapped across worms.
    }
    \label{fig:3}
\end{figure*}

\subsection{Fluctuating potential landscapes underlie the emergence of heavy tails in \emph{C. elegans} foraging}

A possible explanation for the inability of the Langevin Eq.\,\ref{eq:phi2_dot} (or the full model Eq.~\ref{eq:rho_dot}) to capture heavy tails is the existence of subtle \emph{hidden} fluctuations that evolve on timescales comparable to the observation time $T_\text{expt}$. The idea stems from observations that worms slowly adapt their search strategy upon removal of food by lowering their rate of ``pirouettes'' to explore wider areas in search for food \cite{Hills2004,Gray2005,Salvador2014,Calhoun2015,Hums2016,Flavell2020}. A time-evolving rate of pirouettes calls for a non-stationary extension of the model via time-dependent drift and diffusion terms\,:
\begin{equation}\label{eq:phi2_dot_t}
    \dot{\phi_2} = F(\phi_2,t) + \sqrt{2 D(\phi_2,t)}\eta(t)\,,
\end{equation}
which reflect adaptation throughout $T_\text{expt}$.

Time-dependent drift and diffusion coefficients are inferred as described in Appendix A. The resulting potential landscape evolves as shown in Fig.\,\ref{fig:3}(a), validating the hypothesis that worms slowly adapt by increasingly performing runs, in agreement with Refs.~\cite{Hills2004,Gray2005,Salvador2014,Calhoun2015,Hums2016,Flavell2020}. Over time, the ``run-and-pirouette'' random walk is indeed biased to explore further away. Notably, time dependency is sufficient to reproduce heavy tails and long-range correlations exhibited by the worms, Fig.\,\ref{fig:3}(b).

\section{Slowly fluctuating landscapes: generality of heavy tails}

Our data-driven results show that the observed heavy tails result from slow adaptation. Could similar mechanisms more generally underlie the widespread observation of heavy tails across behaving animals? Animals do modulate their behavior, either due to environmental factors or through endogenous  internal states driven by neuromodulation, such as hunger or stress \cite{Bargmann2012,Marder2012,Flavell2022}. Such a continuum of scales inevitably results in non-stationarity since long-lived modes prevent relaxation within a finite observation time $T_\text{expt}$. Our goal here is to investigate theoretically the role of non-stationary fluctuations.

\medskip
\subsection{A fluctuating landscape picture of animal behavior}

Given a set of observations of animal locomotion, we decompose the dynamics into ergodic, $x$, and non-ergodic, $s$, components. The former are the variables that mix rapidly and define the potential wells that correspond to the stereotyped behaviors\,; the latter evolve on timescales $\tau_s\sim T_\text{expt}$ and slowly modulate the potential landscape of $x$. Assuming an appropriate timescale separation, we can describe the long-term dynamics by the following phenomenological model\,:
\begin{align}
    dx_t = -\tau_x^{-1}\partial_x U(x_t,s_t) dt + \sqrt{2T_x\tau_x^{-1}} dW^x_t \label{eq:x},\\ 
    ds_t = -\tau_s^{-1}\partial_s V(s_t) dt + \sqrt{2T_s\tau_s^{-1}}dW^s_t\,\label{eq:s}\,.
\end{align}
By rescaling time we can set $\tau_x = 1$, $dW^x_t$ and $dW^s_t$ are independent increments of a Wiener process, $T_x$ and $T_s$ are the level of fluctuations in $x$ and $s$, $U$ is a potential landscape with multiple wells corresponding to long-lived stereotyped behaviors, and $V$ is uncoupled from the dynamics of $x$ for simplicity. 

In the following sections, we show that the slow modulation of the dynamics introduced by the hidden modes $s$ generally give rise to heavy tails and non-trivial correlations, analogously to the above case of the worm. In addition, we determine the exponent of the tails and show that in a limiting case, it asymptotes to the value $-2$ observed across animal species, see Fig.\,\ref{fig:1}. 

\subsection{Heavy-tailed first passage times in slowly-driven metastable dynamics}

In the context of the Langevin dynamics Eq.\,\ref{eq:x}, the distribution of times spent in a given behavioral state is given by the time  to escape from a potential well, the so-called first passage time distribution (FPTD) \cite{Hanggi1990}, which is of interest across biology, chemistry, finance, physics and mathematics \cite{Szabo1980,Condamin2008,Benichou2014,Chicheportiche2013}. We provide a short pedagogical introduction to first-passage times in Appendix B. Analytical expressions are rare \cite{Grebenkov2015} and most results focus on the mean first passage time (MFPT) \cite{Hanggi1994,Benichou2015}, more tractable but not representative of the long time behavior in the presence of multiple timescales \cite{Godec2016}. To investigate whether the non-ergodic dynamics of Eqs.\,\ref{eq:x},\ref{eq:s} generally yield heavy tails, we derive hereafter the large time asymptotics of its FPTD.

The observation time $T_\text{expt}$ separates ergodic and non-ergodic modes and sets the slowest hopping rate $\omega_\text{min}\sim T_\text{expt}^{-1}$. The long-time behavior of the FPTD out of a static potential well is given by (see Appendix B)
\begin{align}\label{eq:f(t,w)}
    f(t,\omega) = \omega e^{-\omega t}\,,
\end{align}
where the slowest $\omega_\text{min}$ normally dominates the asymptotic behavior. Since the typical time of modulation $\tau_s = O(T_\text{expt})$, we can assume that transitions occur within a nearly static potential, i.e., use Eq.~\ref{eq:f(t,w)} even in the presence of adaptation. However, $\omega_{min}$ will now vary as $s$ fluctuates. The resulting FPTD $f(t)$ is given by the expectation value of $f(t,\omega)$ over the distribution $p(\omega)$ of $\omega$ weighted by the number of transitions within $T_\text{expt}$, which is $\propto \omega$. In short\,:
\begin{equation}\label{eq:f(t)}
    f(t) 
    \propto \int_{\omega_\text{min}}^{\omega_\text{max}} p(\omega)\times \omega \times \omega e^{-\omega t}d\omega.
\end{equation}
The tail of the distribution is dominated by instances where the barrier height is the largest, motivating the use of Kramers approximation \cite{VanKampen1992,Hanggi1990}\,:
\begin{equation}\label{eq:Kramers}
    \omega(s) = \omega_0 \exp\left\{-\frac{\Delta U(s)}{T_x}\right\}\,,
\end{equation}
where $\Delta U(s)$ is the height of the barrier to be overcome and $\omega_0$ is a constant frequency (see Appendix B). Assuming that each measurement starts from initial conditions sampled according to a Boltzmann weight, the distribution of $s$ is given by \footnote{We note that this is generally true even for fixed initial conditions as long as $\tau_s<T_\text{expt}$. If $\tau_s \sim T_\text{expt}$, then $p(\omega)$ is primarily defined by the distribution of initial conditions. However, when the initial conditions are well approximated by a normal distribution with variance $\sigma^2$, the denominator in the Boltzmann weight should be changed accordingly and this will change the final form of the first passage time distribution. Nonetheless, the derivation we present is general and can be adapted for a given $p(s)$, see Appendix C.}
\begin{equation}\label{eq:Boltmzann_s}
    p(s) \propto \exp\left\{-\frac{V(s)}{T_s}\right\}\,.
\end{equation}
When the barrier height fluctuations are large enough to yield $\omega_\text{min}^{-1}\sim T_\text{expt}$, we can combine equations Eqs.\,\ref{eq:f(t)},\ref{eq:Kramers},\ref{eq:Boltmzann_s} to obtain the large $t$ limit of the FPTD\,:
\begin{equation}\label{eq:f(t)_asymp}
    f(t) \sim t^{-2} \exp\left\{-\frac{V(\Delta U^{-1}(T_x \log(\omega_0 t)))}{T_s}\right\}\,.
\end{equation}
Here, $\Delta U^{-1}(\cdot)$ represents the inverse function of $\Delta U(s)$ and we kept only the dominant order of the asymptotic approximation (see Appendix C). Importantly, when $T_s\rightarrow\infty$ we obtain $f(t)\sim t^{-2}$ under very general assumptions on the form of $V(s)$ and $U(x,s)$. In addition, when $V(s)$ and $\Delta U(s)$ are asymptotically equivalent, i.e., grow with the same power of $s$ at large $s$, the distribution $f(t)$ behaves as a power law $f(t)\sim t^{-2-c\frac{T_x}{T_s}}$ with a correction to $-2$ proportional to $\frac{T_x}{T_s}$. In Fig.\,S3 we confirm our theoretical predictions using numerical simulations of a Poisson process with varying hopping rates (see Appendix C). Eq.~\ref{eq:f(t)_asymp} thus shows that slow modulation, which may result from interactions with the environment and/or slowly varying internal states \cite{Bargmann2012,Marder2012}, can indeed generally yield heavy-tailed FPTD.

\begin{figure*}
    \centering
    \includegraphics{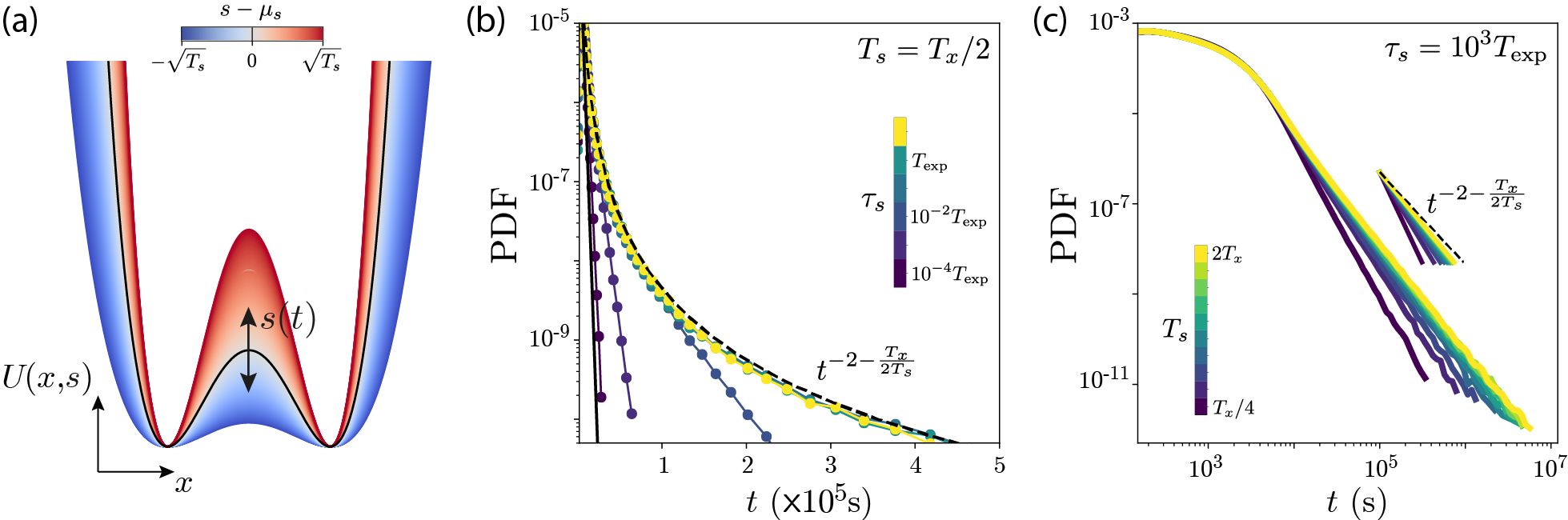}
    \caption{{\bf Emergence of heavy-tails in the first passage time distribution of a slowly-driven double-well potential. }
(a) Schematic of the variation of the double-well potential with $s$ (colored from blue to red;  the black line represents $s = \mu_s$). 
    (b) Probability density function (PDF) of times spent in a potential well (or equivalently, first passage time distribution) from numerical simulations of Eq.\,\ref{eq:(x,s)_DW} for different values of $\tau_s$ and $T_s = T_x/2$ (see Appendix A).  When $\tau_s\rightarrow 0$, the potential landscape relaxes to its mean value faster than the time to escape the well, resulting in an exponential  with a hopping rate corresponding to $\mu_s$ (black line). As $\tau_s$ approaches $T_\text{expt}$, we observe a transition from exponential to power law, and in the limit of large $\tau_s$ we obtain the power law Eq.\,\ref{eq:fpt_dw_asymptotics} (black dashed line).
    (c) Probability density function (PDF) of times spent in a potential well from numerical simulations of Eq.\,\ref{eq:(x,s)_DW} for large $\tau_s = 10^3 T_\text{expt}$ and different values of $T_s$ (see Appendix A). As predicted, the tail of the distribution behaves as $f(t)\sim t^{-2-\frac{T_x}{2T_s}}$ (colored lines) with an exponent that approaches $-2$ as $T_s\rightarrow\infty$ (black dashed line).}
    \label{fig:4}
\end{figure*}

\subsubsection{Slowly-driven double-well potential}

As a further illustration of our result Eq.~\ref{eq:f(t)_asymp}, we consider a double-well potential whose barrier height is slowly modulated according to an Ornstein-Uhlenbeck process, Fig.\,\ref{fig:4}(a). The dynamics of $x$ and $s$ are given by
\begin{equation}\label{eq:(x,s)_DW}
    \begin{cases}
        dx_t =   - 4 s_t^2 x_t (x_t^2-1) dt + \sqrt{2T_x}dW^x_t\,, \\ 
    ds_t =  -\tau_s^{-1} (s_t-\mu_s)dt +\sqrt{2T_s\tau_s^{-1}}dW^s_t\,,
    \end{cases}
\end{equation}
where $T_x=10^{-3}$, $\mu_s = \sqrt{T_x}$, $\tau_s = 10^{3}T_\text{expt}$ (see Appendix A for details).  Since the tail of $f(t)$ is dominated by large $s$ values, we can take $V(s)\sim s^2/2$, and thus $V(\Delta U^{-1}(x)) \sim x/2$. Eq.\,\ref{eq:f(t)_asymp} predicts then
\begin{equation}\label{eq:fpt_dw_asymptotics}
    f(t) \sim t^{-2-\frac{T_x}{2T_s}}\,.
\end{equation}
To test this result we performed direct numerical simulations of Eq.\,\ref{eq:(x,s)_DW} while varying $T_s$ and $\tau_s$. Results in  Figs.\,\ref{fig:4}(b,c) (see also Fig.~S4) quantitatively confirm the dependence of the power law exponent on the ratio $T_x/T_s$, and its approaching $t^{-2}$ as $T_s\rightarrow\infty$.  These results support our theoretical predictions, and provide further intuition for how heavy-tailed distributions depend on the behavioral variability $T_x$, the strength of adaptation $T_s$, and the timescale of behavioral adaptation $\tau_s$.

\subsection{Long-range correlations and finite-size corrections in slowly-driven metastable dynamics}

This Section complements the previous one by showing that slow modulation also induces heavy tails and long-range anti-correlations in the correlation function, as observed for the worm data in Fig.\,\ref{fig:3}(b-right). 

\subsubsection{Heavy tails}

The connected correlation function of $x$ in Eq.~\ref{eq:x} is
\begin{equation}\label{eq:C_x(tau)}
    C_x(\tau) = \frac{\left\langle x(t)x(t+\tau)\right\rangle - \langle x \rangle ^2}{\langle x^2 \rangle - \langle x \rangle^2}\,,
\end{equation}
where $\langle \cdot \rangle$ represents the ensemble average over the invariant density. In a static landscape, the long-time behavior of $C_x$ is dominated by the first non-trivial eigenvalue of the Fokker-Planck operator $\Lambda_2$, which is proportional to the slowest hopping rate $\Lambda_2 \propto \omega_{min}$ \cite{Risken1989,Coffey2004}, i.e., $C_x(\tau) \sim e^{-\Lambda_2 \tau}$. As in the previous section, when the landscape is slowly modulated, $\Lambda_2$ and $\omega_{min}$ fluctuate, and $C_x$ is given by a weighted average over $p(\omega)$\,:
\begin{align}\label{eq:C_x_avg}
    C_x(\tau) \sim \int_{\omega_\text{min}}^{\omega_\text{max}}  p(\omega)\times e^{-\omega \tau} d\omega\,.
\end{align}
Notice that, compared to Eq.\,\ref{eq:f(t)}, the integrand is divided by $\omega^2$: one $\omega$ is dropped since $C_x(t) \approx f(t,\omega)/\omega$ and the other $\omega$ is the number of hoppings, which ought to be counted for $f(t)$ but not for $C_x$. Following the same steps as for $f(t)$ (see Appendix E), we predict
\begin{align}\label{eq:C_x_asymp}
    C_x(\tau) \sim \exp\left\{-\frac{V(\Delta U^{-1}(T_x \log(\omega_0 \tau)))}{T_s}\right\},
\end{align}
to the dominant order for large $\tau$'s. As for the FPTD, when $V(s)$ and $\Delta U(s)$ are asymptotically equivalent, $C_x(\tau)\sim \tau^{-c\frac{T_x}{T_s}}$, with the same constant $c$ as for the FPTD. In this case, $f(t)\sim t^{\beta}$ and $C_x(\tau) \sim \tau^\gamma$ with exponents related by $\gamma = \beta + 2$. 

\begin{figure*}
    \centering
    \includegraphics{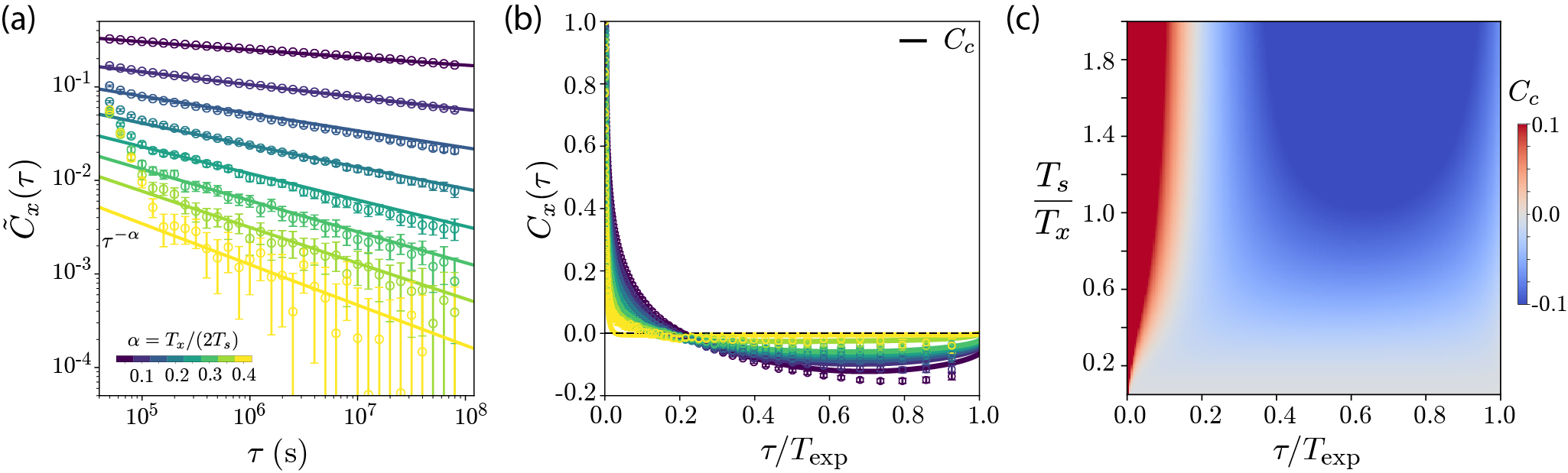}
    \caption{{\bf Power-law correlations and finite-size corrections in a slowly-driven double well potential. }
    (a) Estimated non-connected correlation function $\tilde{C}_x(\tau) = \langle x(t) x(t+\tau)\rangle$ (see Appendix A) for the position $x$ of a particle in a double well potential driven on a timescale $\tau_s = 10^2 T_\text{expt}$. Our prediction $C_x(\tau)\sim \tau^{-\frac{T_x}{2T_s}}$ is validated (solid lines). Error bars represent 95\% confidence intervals across 50,000 simulations. 
    (b) Connected autocorrelation function, $C_x(\tau)$, directly estimated through time averages (see Appendix A), for a double well potential driven on a timescale $\tau_s = 10^2 T_\text{expt}$. Due to the existence of timescales $\sim T_\text{expt}$, finite-size corrections $C_c$ are present and generate long-range anti-correlations in $C_x(\tau)$, as predicted in Appendix F (solid lines). For both $C_x(\tau)$ and $C_c(\tau)$ we normalize the correlation functions by dividing by their value at $\tau = 1\,\text{lag} = 5\times 10^{-4}T_\text{expt}$. Error bars represent 95\% confidence intervals across 50,000 simulations.
    (c) Finite-size correction $C_c(\tau)$ {\it vs} $T_s$ (see Appendix F). As we increase $T_s$, the range of observed $\omega$ grows, and so do finite-size corrections, which result in stronger anti-correlations (blue). Conversely, for small $T_s$ finite size effects are negligible as the longest sampled $\omega^{-1}\ll T_\text{expt}$.}
    \label{fig:5}
\end{figure*}

To illustrate these results, we return to the double-well potential Eq.\ref{eq:(x,s)_DW}. The expectation would be $C_x(\tau) \sim \tau^{-\frac{T_x}{2T_s}}$ and $\langle x\rangle = 0$ (since the potential is symmetric). Indeed, if we measure the non-connected correlation function from numerical simulations of Eq.\,\ref{eq:(x,s)_DW} (without subtracting the mean, see Appendix A), we recover the theoretical expectation of power-law correlations for large $\tau_s$, Fig.\,\ref{fig:5}(a). 

\subsubsection{Finite-size effects and anti-correlations}
In the previous double-well example, we used symmetry of the potential to avoid estimating the means. In general, we have to resort to empirical estimations via averaging in time (see Appendix A). In the presence of slow time scales, we expect that temporal averages $\hat{\mu}_x = \frac{1}{T_\text{expt}}\int_0^{T_\text{expt}} x(t) dt$ will deviate from ensemble averages, resulting in finite-size effects \cite{Cavagna2018}. In particular, we expect that on average $x(t)-\hat{\mu}_x$ will change sign as time progresses. This transient behavior results in apparent long-range anti-correlations, since $x(t)-\hat{\mu}_x$ and $x(t+\tau)-\hat{\mu}_x$ will tend to have different signs for large $\tau$ \footnote{A similar observation has been done in the analysis of spatial correlation functions in flocks of birds \cite{Cavagna2010}.}. Therefore, we expect that Eq.\,\ref{eq:C_x_asymp} will deviate from empirical estimations when slow time scales are present. Indeed, when we estimate $C_x$ through temporal averages (see Appendix A), we do observe long-range anti-correlations, see Fig.\,\ref{fig:5}(b). Importantly, using our analytical derivation of the non-connected correlation $\hat{C}_x(\tau)$ and results in Ref.~\cite{Desponds2016}, we derive an expression for the finite-size correction $C_c(\tau)$ (see Appendix F), that captures empirical estimate of $C_x(\tau)$, as shown in Fig.\,\ref{fig:5}(b). 

Our predicted dependence of finite size effects on $T_s$ is well confirmed by results in Fig.\,\ref{fig:5}. Furthermore, small $\tau_s$ yield the expected exponential tails for the non-connected correlation function, which become a power law when $\tau_s \sim T_\text{expt}$, see Fig.\,S5(a-left). Note that finite-size corrections are apparent even in the exponential regime as long as its timescale is $\sim T_\text{expt}$, see Fig.\,S5(a-right,b). This corresponds to barriers sufficiently high to produce hopping rates comparable to $T_\text{expt}$. 

To conclude, note that our predictions recapitulate and rationalize the phenomenology of the correlations that we presented in Fig.\,\ref{fig:3}(b-right) for foraging worms.

\section{Testing our theoretical predictions}
\label{sec:test}

Our predictions in the previous section were motivated by the particular example of the wild type nematode {\it C. elegans} but they emerged to be quite general. To challenge this predicted generality, we decided to consider different animal examples\,: a genetic mutant of \emph{C. elegans} and the larval zebrafish. Both turn out to agree with our predictions and validate the idea that behavioral plasticity is a minimal, yet necessary ingredient for the emergence of heavy tails in animal behavior.

\subsection{A mutation in the \emph{C. elegans} \emph{npr-1} gene suppresses heavy tails observed in wild type animals}

\begin{figure*}
    \centering
    \includegraphics{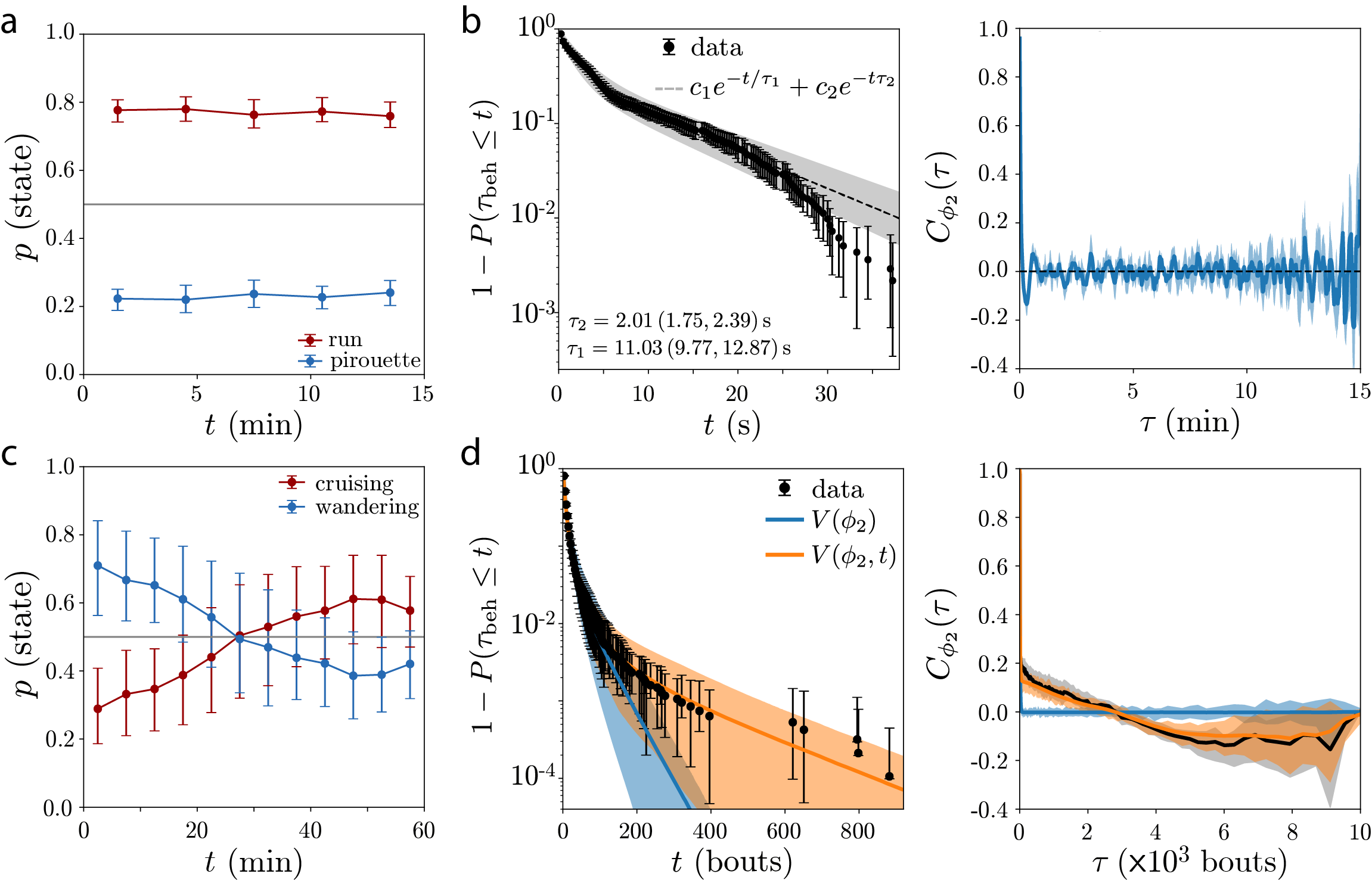}
    \caption{{\bf Testing our theoretical predictions: loss-of-function mutation in the \emph{npr-1} gene of \emph{C. elegans} ablates heavy-tailed statistics, and larval zebrafish exhibit heavy-tails in their long-lived behavior.}
    (a) Probability of observing ``run'' and ``pirouette'' states as a function of time for \emph{npr-1} mutants. Unlike wild type N2 worms, the distribution of these states is constant over time. We estimate the fraction of time spent either performing a ``run'' or a ``pirouette'' in $3\,\text{min}$ long windows, and obtained bootstrapped averages across all 7 worms.
    (b-left) Complementary cumulative distribution function of observing a ``run'' or a ``pirouette'' with a duration $>\tau_\text{beh}$, $1-P(\tau_\text{beh}\leq t)$ (same as Fig.2d-left). The distribution is well approximated by a sum of exponential functions: a longer one that corresponds to the ``runs'' and a shorter one that corresponds to the ``pirouettes''. Unlike N2 worms, the first passage time distribution in \emph{npr-1} mutants is not heavy-tailed.
    (b-right) Autocorrelation function $C_{\phi_2}$ of the slow mode $\phi_2$ inferred from the \emph{npr-1} mutant data. Unlike wild type N2 worms, non-trivial long-range correlations are absent and instead correlations decay within a minute.
    (c)  We collected data from \cite{marques2018structure} in which larval zebrafish are exposed to a chasing dot stimulus for $5\,\text{s}$ every $2\,\text{min}$ for at least 1 hour. We proceed as for \emph{C. elegans} (see Appendix A), and find that the longest-lived dynamics $\phi_2$ corresponds to a ``wandering-cruising'' axis, in which the fish either engages in bout sequences with large orientation changes (``wandering''), or performs sequences of smoother forward bouts (``cruising''), Fig.\,\ref{fig:S_npr-1_zebrafish}(b). Over time, we observe that the probability of the ``cruising'' state slowly increases over time, becoming more prevalent than the ``wandering'' state. We estimate the probability of being in the ``cruising'' and ``wandering'' states in $5\,\text{min}$ windows, and show the average probability bootstrapped across the 11 fish.
    (d-left) Complementary cumulative distribution function of observing ``cruising'' or ``wandering'' with a duration $>\tau_\text{beh}$, $1-P(\tau_\text{beh}\leq t)$ (same as Fig.2d-left). We plot the data in black, as well as the distribution obtained from simulations with a stationary model ($V(\phi_2)$, blue)  and a time-dependent model ($V(\phi_2,t)$, orange), see Appendix A for details. As for \emph{C. elegans} N2 worms foraging, we find that time-dependent parameters are required to accurately predict the tail of the distribution.
    (d-right) Autocorrelation function $C_{\phi_2}$ of the slow mode dynamics obtained from the larval zebrafish data (black) as well as simulations from the stationary model (blue) and the time-dependent model (orange). Notably, we find that larval zebrafish exhibit heavy-tailed statistics due to the explicit time-dependency of the behavioral dynamics.
    }
    \label{fig:6}
\end{figure*}

The NPR-1 neuropeptide receptor is known to impact several \emph{C. elegans} behaviors, viz., aerotaxis and food response \cite{Bargmann2012}. We collected a public dataset where worms of the \emph{npr-1} loss-of-function strain \emph{npr-1(ad609)} are allowed to freely explore an agar plate with a uniform food patch (see Appendix A) and used the same method as in Fig.\,\ref{fig:2} for wild type worms. The upshot is that the short-time behavior of \emph{npr-1} mutants on food is similar to wild type N2 worms off food. The structure of the behavioral landscape is similar to the one found in wild type \cite{Costa2023markovian}, with the dominating kinetics being the transitions between ``runs'' and ``pirouettes'', see Fig.\,S7(a). However, mutants and wild type crucially differ at long timescales as mutants do not exhibit heavy tails, see Fig.\,\ref{fig:6}(b). Instead, the tail of the first passage time is close to an exponential, and correlations decay to zero within a minute. This is due to the mutants' inability to adapt their pirouette rates over time, Fig.\,\ref{fig:6}(a), contrary to the modulation highlighted in Fig.~\ref{fig:3}(a) for the wild type.

\subsection{Heavy tails in the behavior of larval zebrafish results from slow habituation}

Zebrafish larvae move in discrete tail bursts, interspersed by periods of immobility. We leverage a dataset previously analyzed in \cite{marques2018structure,Groneberg2020}, where larvae are exposed to a ``chasing dot'' stimulus for $5\,\text{s}$ every $2\,\text{mins}$ for at least one hour (see Appendix A for details). The dataset consists of a sequence of bouts, in which the curvature of the tail of the fish is tracked at a high spatiotemporal resolution yielding a time series of cumulative tail angles for each bout, see Fig.\,S7(b). 

From these bout sequences, we proceed as for the postures of \emph{C. elegans}, and find that $K^*=5\,\text{bouts}$ and $\tau^*=3\,\text{bouts}$ yield an accurate description of the long-lived dynamics (see Appendix A for details). Using these parameters, we find that the dominating long-lived mode $\phi_2$ captures transitions between sequences of smooth forward bouts (``cruising''), and sequences of sharp orientation changes (``wandering''), as shown in Fig.\,S7(b). Projecting the full dynamics onto $\phi_2$ yields a time series along this ``cruising-wandering'' axis that fluctuates over time.

Over long timescales, we observe that fish modulate time spent in the cruising state, Fig.\,\ref{fig:6}(c), likely due to the habituation to the stimulus condition. That is confirmed by the observation that, as in \emph{C. elegans}, the inferred potential landscape is slowly varying in time, and this slow modulation is essential to accurately predict the heavy-tailed FPTD and the non-trivial long-range correlations shown in Fig.\,\ref{fig:6}(d).

\section{Discussion}

The combination of theory, numerics, and experimental data analysis we used here provided evidence that the multiplicity of timescales inherent to animal behavior is sufficient to give rise to heavy-tailed first passage times and long-range correlations. The phenomenon is demonstrated in Fig.~\ref{fig:2}(d), which shows first passage time distributions (FPTD) and correlation functions for the \emph{C. elegans} nematode. We started from this example as high-resolution measurements of the animal pose are available, bridging from $\sim 0.1\,\text{s}$ chaotic posture dynamics to $\sim 10\,\text{s}$ stochastic hopping among ``runs-and-pirouettes'' \cite{Fujiwara2002}. 

To capture long-term effects, we combined ideas from reduced order modeling \cite{Givon2004,Coifman2008,Froyland2014,Giannakis2019} and stochastic model inference \cite{Lamouroux2009,Callaham2021,Frishman2020}. The resulting one-dimensional stochastic differential equation yields an overdamped description of a partially-observed system with metastable dynamics. Our contribution here is mostly methodological\,: rather than assuming structure \emph{a priori}, we aim for a coarse-grained simplified description and let data drive the process of building it. While already effective, each step in our analysis can likely be enhanced by modern tools from machine learning (see, e.g., \cite{Dietrich2023}), an issue that we leave for future work.

The static version of our coarse-grained description is unable to capture heavy tails and long-range correlations displayed by experimental data. This failure is not the result of our approximation as a full, yet still autonomous, model of the dynamics is also missing these effects. Both are useful to capture short-term properties, but some new ingredient is needed to capture long times. The idea that we pursued here is that non-ergodic modulations, varying over timescales comparable to the observation time $T_\text{expt}$, are the key. In practice, this amounts to having an effective potential landscape slowly changing in time. Fig.\,\ref{fig:3} demonstrated that long-term modulation allows us to capture the missing effects exhibited by the worm. In addition, we find that the slow modulation reflects the adaptation of the worm's foraging strategy to searching for food further away. In this way, we go beyond classical approaches to reduced-order modeling, recognizing the existence of non-ergodic fluctuations and introducing a non-stationary model to encode them.

Our analysis makes strong predictions that can be tested experimentally. In particular, perturbing the neural mechanisms responsible for the adaptation of ``pirouette'' rates, such as dopaminergic and glutamatergic signaling \cite{Hills2004}, should alter the long-term features of behavior. We predict that, in the absence of adaptation, shorter timescale movements remain unaffected while the dwell times in the ``run'' and ``pirouette'' states would become exponentially distributed, rather than heavy-tailed, and that the correlation function would simply decay exponentially to zero on fast timescales. Mutants of the NPR-1 neuropeptide receptor that we analyzed in Section~\ref{sec:test} are our first, promising and positive step in that direction. New mutants and data would be important to further confirm our predictions. It would also be interesting to check whether wild strains of \emph{C. elegans} exhibit a similar pattern as the one found here for the laboratory strain N2. Indeed, laboratory strains of \emph{C. elegans} were grown in relatively poor conditions for multiple generations, when compared to much richer natural environments encountered by wild strains \cite{Felix2010,Frezal2015,Schulenburg2017}. This has led to an evolutionary divergence between the laboratory strain N2 and wild strains, with N2 worms fixing several mutations that affect a variety of phenotypes \cite{McGrath2009,Green2014,Andersen2014,Sterken2015}.  Whether the observed heavy tails are also observed in wild strains, and how they are modulated depending on the natural habitat of different worm species, would provide further insight into the ecological and evolutionary significance of heavy tails. We expect that richer environments yield behavioral modulations on a wider range of time scales and the exponent of the heavy tails will reflect the natural habitat of different species \cite{Beets2020}. The explicit formulae Eqs.~\ref{eq:f(t)_asymp} and \ref{eq:C_x_asymp} that we derived here should enlighten the contributions stemming from variations in the potential landscape (encoded in $T_s$), fluctuations in each behavior (encoded in $T_x$) and timescale of adaptation (encoded in $\tau_s$). As for our predictions on long-range correlations, recent work that followed the original arXiv version of this manuscript, provided evidence for power-law correlations in fruit flies \cite{Bialek2023_PRL}. Their observations are consistent with our theoretical predictions, and we argue that they might stem from non-ergodic internal states. Indeed, we expect there to be slow modes that evolve on timescales comparable to the  $1\,\text{hour}$ recordings used in \cite{Bialek2023_PRL}, see \cite{Qiao2018,Overman2022}. Future work will be needed to shed light on this issue. 

Heavy-tailed distributions in the duration $t$ of behaviors with an exponent $f(t)\approx t^{-2}$ are found across multiple species, from bacteria \cite{Korobkova2004}, termites \cite{Miramontes2014} and rats \cite{Jung2014} to marine animals \cite{Humphries2010,Sims2008}, humans \cite{Raichlen2014} and even fossil records \cite{Sims2014} (see Fig.\,\ref{fig:1}). Such observations have led researchers to hypothesize that L\'evy flights yield optimal search strategies when the power law exponent is $-2$ \cite{Viswanathan1999,Wosniack2015,Wosniack2017,Guinard2021,Clementi2021}, although this view has been met with some controversy \cite{Pyke2015,Reynolds2015}. Here, we provided evidence in two distant model organisms, {\it C. elegans} and zebrafish, that such fat tails may simply emerge from modulation over time and thus be a by-product of the evolutionary favorable ability to perform adaptive behavior \cite{Mery2010}.

Power laws are observed beyond behavior, from solar flares \cite{Wheatland1998,Boffetta1999} to the brain \cite{Beggs2003} and the idea of variable barrier heights has appeared in different contexts (for a review, see, e.g., \cite{Newman2005}). In disordered systems, averaging over an exponential distribution of barrier heights can give rise to a broad distribution of waiting times \cite{Bouchaud1990,Ben-Avraham2000}. Note that, while this mechanism is qualitatively analogous to the one presented here, ours relies on the temporal (rather than spatial) variation of barrier heights. That results in distinct emergent behavior that depends directly on the measurement time scale $T_\text{expt}$ (that sets the lowest hopping rate $\omega_\text{min}$) and the magnitude of the non-ergodic fluctuations. Time-dependent energy barriers have also been used in bacterial chemotaxis \cite{Tu2005}. The analysis of \cite{Tu2005} concerns a particular limit of our derivation, in which $T_s\rightarrow\infty$ and the distribution of hopping rates becomes uniform. Our derivation considered a more general dynamics, and predicted corrections to the power laws that go beyond the limits in \cite{Tu2005}. Another proposal is the presence of multiplicative noise terms in the dynamics \cite{Biro2005,Lubashevsky2009}, and this notion has recently been used to explain the emergence of L\'evy flights in the collective behavior of midge swarms \cite{Reynolds2016}. Our Eqs.\,\ref{eq:x},\ref{eq:s} do give rise to an effectively colored multiplicative noise term for the quasi-stationary behavioral dynamics but we go beyond by determining the dependency on the relationship between the correlation time of the colored noise $\tau_s$ and the measurement time $T_\text{expt}$, and between the additive and multiplicative noise terms. Finally, some of the arguments we have put forward have appeared in discussions of ``criticality'' \cite{Touboul2017,Priesemann2018,Morrell2023}. That is the apparent tendency of some systems to sit between two qualitatively different ``phases'' (see, e.g., \cite{Mora2011}), which makes them akin to critical systems in statistical mechanics \cite{Cocchi2017,OByrne2022}. Our derivations here apply to a wider range of model classes, using the framework of out-of-equilibrium statistical mechanics to explicitly connect the long time scale emergent behavior with the underlying effective fluctuations. In addition, unlike other approaches \cite{Schwab2014,Priesemann2018}, our framework does not require explicit external drives, but simply collective modes that evolve in a weakly non-ergodic fashion.

On the theoretical side, to derive the analytical expressions~\ref{eq:f(t)_asymp} and \ref{eq:C_x_asymp} of correlation function and FPTD, we exploited a separation between microscopic dynamics and long-time behavior. Further work will be required when such separation and the quasi-adiabatic approximation do not hold. For example, we find numerically that for intermediate values of $1\ll \tau_s \ll T_\text{expt}$ and finite $T_s$, the FPTD behaves as a truncated power law with an exponent $>-2$. In this regime, the barrier heights fluctuate significantly before the particle hops. Intuitively, we expect that if barrier-crossing events become uncorrelated, the $\omega$ factor in the FPTD that accounts for the number of hopping events drops out, resulting in an exponent $-1$ rather than $-2$ \footnote{Interestingly, fast-fluctuating hopping rates and scale-invariance arguments have been used to explain heavy-tailed distributions of uncorrelated resting times in mice \cite{Proekt2012}}. In the opposite limit, when $\tau_s \gg T_\text{expt}$, it is the distribution of initial conditions (which we took to be Boltzmann) that determines the emergent behavior. This assumption holds if we consider that behavioral ``individuality'' is equivalent to having a very slow mode $\tau_s \gg T_\text{expt}$. This would mean that different animals in a population of conspecifics will exhibit a degree of ``individuality'' that matches the steady-state distribution of such long-lived modes. Intriguingly, such a relationship between inter-individual variability and long-lived behavioral variability is observed in flies \cite{Hernandez2021}. In this sense, when $\tau_s \gg T_\text{expt}$, our results are equivalent to explaining the emergence of heavy tails through inter-individual variability \cite{Petrovksii2011}. If such variability differs from the Boltzmann assumption, heavy tails need to be corrected accordingly, following the steps of our derivation but with a modified $p(s)$.

In conclusion, we have used a physics approach to shed light on animal behavior, leveraging statistical mechanics as a framework for thinking about the effect of slowly-varying modulation, either environmental or by internal states. Concurrently, observations from animal behavior inspired new physical results regarding the emergence of heavy tails in slowly-driven potential landscapes, which are generally relevant to a wide range of fields in chemistry, biology, or finance (see, e.g., \cite{Szabo1980,Condamin2008,Benichou2014,Benichou2015,Ghusinga2017,Chicheportiche2013,deWit2022} and references therein).

\section{Acknowledgements}

We thank Adrian van Kan, St\'{e}phan Fauve, Federica Ferretti, Tosif Ahamed, Nicola Rigolli and Arghyadip Mukherjee for their comments. We also thank Antonia Groneberg for sharing the dataset and João Marques for providing insight on larval zebrafish behavior. This work was partially supported by the LabEx ENS-ICFP: ANR-10-LABX-0010/ANR-10-IDEX-0001-02 PSL* and by the NIH Grant No. 1RF1NS128865-01. A.C. also acknowledges useful discussions at the Aspen Center for Physics, which is supported by the National Science Foundation, Grant No. PHY-1607611. G.S. was supported by the European Union’s Horizon 2020 Research and Innovation program under the Marie Skłodowska-Curie Grant No. 813457.

\setcounter{equation}{0}

\renewcommand{\theequation}{A\arabic{equation}}

\section*{Appendix A: Methods}

\noindent{\bf Software and data availability:} Code for reproducing our results is publicly available:
\url{https://github.com/AntonioCCosta/fluctuating_potential}. Data can be found in \cite{manuscript_data}. 

\medskip

\noindent{\bf \emph{C. elegans} foraging dataset: } We used a previously-analyzed dataset \cite{Stephens2008}, in which young-adult N2-strain \textit{C.~elegans} were tracked at $f=16\,{\rm Hz}$ \cite{Broekmans2016}. Worms were grown at $20{^\circ C}$ under standard conditions \cite{Sulston1974}. Before imaging, worms were removed from bacteria-strewn agar plates using a platinum worm pick, and rinsed from \textit{E.~coli} by letting them swim for $1\, {\rm min}$ in NGM buffer. They were then transferred to an assay plate ($9\, {\rm cm}$ Petri dish) that contained a copper ring ($5.1\, \rm{cm}$ inner diameter) pressed into the agar surface, preventing the worm from reaching the side of the plate. Recording started approximately $6\, \rm{min}$ after the transfer and lasted for $T_\text{expt} = 35\, \rm{mins}$. 

\medskip

\noindent{\bf Data-driven reduced order model of \emph{C. elegans} foraging dynamics:} Building upon previous work \cite{Ahamed2021,Costa2023markovian,Costa2023}, we extract a slow reaction coordinate that captures transitions between ``runs'' and ``pirouettes'' from the posture dynamics of \emph{C. elegans}. As in \cite{Broekmans2016}, we extract the centerline of the animal at every $\delta t = 1/16\,\text{s}$, estimate 100 tangent angles along the body, $\theta_t$, and project these angles onto an eigenworm representation \cite{Stephens2008} to obtain a 5-dimensional $35\,\text{minute-long}$ time series, $\theta_t\rightarrow \vec{a}_t \in \mathbb{R}^5, t\in \{\delta t,2\,\delta t, \ldots, 33600\,\delta t\}$, that accurately captures the changing postures of each of the 12 animals (a combined total of 403,200 frames). From such measurements, we perform a time-delay embedding \cite{Costa2023markovian,Ahamed2021} to include short-term memory into an expanded maximally predictive state $X_{K^*}(t) = \vec{a}_{t-K^*:t}$, where $\vec{a}_{t-K^*:t} = \{\vec{a}_{t-K^*},\ldots,\vec{a}_{t-1}\}$. The amount of time delays $K^*$ used to reconstruct the state space is chosen so as to maximize predictive information \cite{Costa2023markovian,Costa2023}. In this way, all the dynamics that mix on a sufficiently fast timescale (compared to the measurement time $T_\text{expt}$) should be included in the state. We note that, as evidenced throughout the manuscript, slow modes that evolve of timescales comparable to the measurement time do not provide enough independent samples to be directly inferred. Such non-stationary dynamics cannot be accounted for by a stationary model, and are thus not recoverable with a delay embedding due to their statistical insignificance. As in \cite{Costa2023markovian}, we choose $K^*=11\,\text{frames} = 0.6875\,\text{s}$ to maximize predictive information within the limits of the data. We then partition the state space into a large number of discrete symbols through k-means clustering, and choose the number of partitions, $N^*=1000$, so as to preserve as much information as possible in the discretization, while avoiding finite-size effects (as in \cite{Costa2023markovian}). The outcome of the partitioning is a symbolic sequence, where each symbol $s_i$ corresponds to a small region of the state space (a collection of similar posture sequences). We then build a Markov chain by counting transitions among state-space partitions separated by a timescale $\tau$, $P_{ij}(\tau) = P(s_j(t+\tau)|s_i(t))\approx e^{\mathcal{L}\tau}$ \cite{Costa2023markovian,Costa2023}, effectively approximating the action of the Perron-Frobenius operator (see, e.g., \cite{Bollt2013}). The transition time $\tau^*=0.75\,\text{s}$ was chosen so as to self-consistently capture the long-lived dynamics \cite{Costa2023markovian,Costa2023}. The eigenfunctions of the Perron-Frobenius operator, and its adjoint, the Koopman operator, capture global patterns of the dynamics that relax to the steady-state distribution on different timescales. In particular, the slowest eigenfunctions of these operators offer optimal reaction coordinates that capture the slow dynamics of the system \cite{Coifman2008,Bittracher2018,Giannakis2019}. The slowest left eigenvector of the reversibilized transition matrix $\phi_2$ captures transitions among ``runs'' and ``pirouettes'' that \emph{C. elegans} uses to forage \cite{Costa2023markovian}. We find the transition point $\phi_2^c$ between ``runs'' and ``pirouettes'' by maximizing the overall coherence of the metastable states \cite{Costa2023markovian,Costa2023}, and recenter and rescale $\phi_2$ to have $\phi_2^c=0$ at the transition point and to have equally-spaced values within $[-2,\phi_2^c=0]$ and $[\phi_2^c=0,2]$ \cite{Froyland2014}. Finally, each symbol $s_i$ assumes a particular value of $\phi_2$, and so we can translate the symbolic sequence into a stochastic time series $\phi_2(t)$ that captures transitions between ``runs'' and ``pirouettes'', see Fig.\,\ref{fig:2}(a-c).

\medskip

\noindent{\bf \emph{npr-1} dataset: } We used a publicly available dataset collected  by the Schafer lab \cite{Yemini2013}, in which young-adults from the \textit{C.~elegans} \emph{npr-1(ad609)} mutant strain were imaged for 15\,\text{minutes} in food-rich plates. Worms were grown at $22{^\circ C}$ under standard conditions, and adult worms were transferred to a 3.5\,\text{cm} diameter plate with a nearly circular bacterial food lawn consisting of 20$\,\mu\text{l}$ of OP50 (see \cite{Yemini2013} for experimental details). After waiting $30\,\text{mins}$ for habituation, worms are then imaged for a total of $15\,\text{mins}$. We analyzed the videos using {\tt wormpose} \cite{Hebert2021} to ambiguous poses when the worm is coiled and obtain a continuous time series of postures. From the full set of 12 worms analysed with {\tt wormpose}, we kept only those that were successfully tracked for at least $80\,\%$ of the frames and for which we could unambiguously assign a dorsal-ventral axis (through the bias of the $a_3$ distribution as in \cite{Hebert2021}). This resulted in a subset of 7 worms that are fully tracked at $20\,\text{Hz}$ for a combined total of 126,000 frames (18,000 per worm). We analyzed these data in the same way as the N2 wild type worms: we project the posture onto an ``eigenworm'' space obtaining $\vec{a}_t \in \mathbb{R}^5, t\in \{\delta t,2\,\delta t, \ldots, 18000\,\delta t\}$, and extract the long-lived dynamics $\phi_2$ using $K^* = 0.4\,\text{s}$, $N^* = \text{int}(10^{2.75}) = 562$ partitions and $\tau^*=0.5\,\text{s}$. These parameters are chosen using the same criteria as for wild type N2 worms: the full analysis pipeline can be found at \url{https://github.com/AntonioCCosta/fluctuating_potential}. We made the time series of eigenworm coefficients for the \emph{npr-1} worms publicly available at \cite{manuscript_data}.

\medskip

\noindent{\bf Zebrafish (\emph{Danio rerio}) dataset: } We use a dataset that was previously analysed in \cite{Groneberg2020}. See \cite{marques2018structure,Groneberg2020} for the experimental details. In these data, wild type T\"{u}bingen fish (6-7 days post fertilization) are exposed to a ``chasing dot'' stimulus for 5 seconds every 2 minutes for at least one hour. The ``chasing dot'' stimulus consists of a dark spot with $1\,\text{mm}$ radius that starts $2\,\text{cm}$ away from the fish and approaches it from one of four directions (left, right, forward, back). Video tracking was performed using custom methods (see \cite{marques2018structure}) that extract kinematic parameters from each burst of tail motion (bout). In our dataset, each bout is represented by 8 cumulative tail angles (measured along 9 points from the swim bladder to the tip of the tail) over time for $250\,\text{ms}$, which corresponds to $175\,\text{frames}$ sampled at $700\,\text{Hz}$. We project each bout, which is effectively a point in a $175\,\text{frames} \times 8\,\text{tail angles}$ dimensional space, to a 20 dimensional space using Principal Component Analysis, preserving virtually all the significant information in the bout space (when compared to a shuffle). To gather enough long timescale data, we selected recordings with at least $7500\,\text{bouts}$, yielding a total of 11 fish recordings and a combined total of 107,260 bouts. We have made the time series of bouts publicly available at \cite{manuscript_data}, and the analysis pipeline can be found at \url{https://github.com/AntonioCCosta/fluctuating_potential}. To extract the slow mode dynamics, we proceeded as we did with the  \emph{C. elegans} data: we reconstruct the long-lived dynamics $\phi_2$ using $K^*=5\,\text{bouts}$, $\tau^*=3\,\text{bouts}$ and $N^* = \text{int}(10^{2.75}) = 562$ partitions, determining the parameters using the same approach as for the other datasets \cite{Costa2023markovian,Costa2023}. 

\medskip

\noindent{\bf Two-dimensional UMAP embedding of the reconstructed state space:} We use the UMAP embedding \cite{McInnes2018} as a tool to visualize the maximally predictive states $X_{K^*}$ of \emph{C. elegans} posture dynamics \cite{Costa2023markovian}. In a nutshell, the UMAP algorithm searches for a low-dimensional representation of the data that preserves its topological structure. We use a publicly available implementation of the algorithm \href{https://github.com/lmcinnes/umap}{{\tt https://github.com/lmcinnes/umap}}, using Chebyshev distances. For wild type N2 worms we use {\tt n\_neighbors}=50 nearest neighbors and {\tt min\_dist}=0.05 as the minimum distance, while for the \emph{npr-1} mutant data we use {\tt n\_neighbors}=15 and {\tt min\_dist}=0.05 reflecting the smaller dataset size.

\medskip

\noindent{\bf Stochastic model inference:} The Kramers-Moyal expansion transforms the master equation for the dynamics of Eq.\,\ref{eq:phi2_dot} into a Fokker-Planck equation

\begin{align}\label{eq:FP_phi_2}
    \partial_t \rho & = \mathcal{L}\rho = \partial_{\phi_2} J(\rho,\phi_2) \nonumber \\ & =-\partial_{\phi_2}\left[F(\phi_2)\rho\right] +  \partial^2_{\phi_2}\left[D(\phi_2) \rho\right],
\end{align}
where $J(\rho,\phi_2)$ is the current, and
\begin{align*}
    F(x) &= \lim_{\tau\rightarrow 0} \frac{1}{\tau} \left\langle \phi_2(t+\tau) - \phi_2(t) | \phi_2(t)=x\right\rangle, \\
    D(x) &= \lim_{\tau\rightarrow 0} \frac{1}{2\tau} \left\langle (\phi_2(t+\tau) - \phi_2(t))^2 | \phi_2(t)=x\right\rangle.
\end{align*}
We use this expansion to estimate $F(\phi_2)$ and $D(\phi_2)$. In practice, given a time series $Y_t$, we estimate the averages in the Kramers-Moyal expansion using a kernel approach \cite{Lamouroux2009},
\begin{align*}
    F(y)_{\tau,h} & = \frac{1}{\tau} \left\langle \frac{K_h(y-Y_t)}{\langle K_h(y-Y_{t^\prime})\rangle_{t^\prime}} (Y_{t+\tau}-Y_t) \right\rangle_t \\ 
    D(y)_{\tau,h} & = \frac{1}{2\tau} \left\langle \frac{K_h(y-Y_t)}{\langle K_h(y-Y_{t^\prime})\rangle_{t^\prime}} (Y_{t+\tau}-Y_t)^2 \right\rangle_t,
\end{align*}
where $K_h(z) = h^{-1}\kappa(z/h)$ and $\kappa$ is the Epanechnikov kernel \cite{Epanechnikov1969,Hardle2006},
\begin{align*}
    \kappa(y) = \begin{cases}
        \frac{3}{4\sqrt{5}}\left(1-\frac{y^2}{5}\right),\, \text{if } y^2<5 \\ 
        0,\, \text{if } y^2>5
    \end{cases}.
\end{align*}
Importantly, the estimator has an explicit dependence on the time delay $\tau$ and the bandwidth $h$. First, as discussed in the main text, we choose $\tau$ long enough such that most of the temporal correlations in the noise have decayed to zero. For \emph{C. elegans} N2 worms crawling off food, it has been shown that $\tau^*=0.75\,\text{s}$ gives an accurate first-order Markov model of the worm dynamics \cite{Costa2023markovian}, and accordingly we find that a stochastic model inferred with $\tau^*=0.75\,\text{s}$ yields nearly delta-correlated noise, Fig.\,S1(b). For the zebrafish dataset, we use the same criterion to choose $\tau^*=3\,\text{bouts}$. Given this time delay $\tau^*$, we choose the bandwidth through the $\Delta$-algorithm introduced in \cite{Lamouroux2009}. In essence, for each bandwidth $h$ we estimate $F_{\tau^*,h}$ and $D_{\tau^*,h}$ and generate simulations with the estimated $F_{\tau^*,h}$ and $D_{\tau^*,h}$. From such simulations, we then re-infer the drift and diffusion from the simulated time series, obtaining $\hat{F}_{\tau^*,h}$ and $\hat{D}_{\tau^*,h}$. Finally, we compare the re-inferred drift and diffusion to the ones estimated directly from the time series,
\begin{align}\label{eq:rec_error}
    \xi (h) = \frac{\int |f_{\tau^*,h}-\hat{f}_{\tau^*,h}|\sqrt{\pi(y) \hat{\pi}(y)}dy}{\int \sqrt{\pi(y) \hat{\pi}(y)} dy},
\end{align}
where $f$ can be either $F$ or $D$, $\pi(y)$ is the steady-state distribution obtained from $F_{\tau^*,h}$ and $D_{\tau^*,h}$ and $\hat{\pi}(y)$ is the one obtained from $\hat{F}_{\tau^*,h}$ and $\hat{D}_{\tau^*,h}$. We choose $h^*$ as the first minimum of $\xi(h)$ \cite{Lamouroux2009}, locally minimizing the difference between original and reconstructed drift and diffusion coefficients and avoiding the trivial minimum that corresponds to $h\rightarrow\infty$ (which yields constant $F$ and $D$). For the \emph{C. elegans} wild type data of Fig.\,\ref{fig:3}, we estimated the change in $\xi (h)$ as a function of $h$, Fig.\,S1(a), which reaches zero at around $h^*\approx 0.1$. We choose $h^*=0.08$ to infer the time series of $\phi_2(t)$. For the zebrafish dataset, we use the same criterion to choose $h^*=0.05$.

\medskip

\noindent{\bf Non-stationary stochastic model inference:} We proceed as before, but now infer  $F_{\tau^*,h^*}^w$ and $D_{\tau^*,h^*}^w$ in overlapping windows. For the \emph{C. elegans} wild type dataset we choose windows of length $5\,\text{min}$ sampled every $30\,\text{s}$, while for zebrafish the overlapping windows were defined using $1000\,\text{bouts}$ sampled every $100\,\text{bouts}$. The window length was chosen long enough to allow for equilibration of the long-lived dynamics, but also short enough such that the steady-state distribution remains approximately constant. 

\medskip

\noindent{\bf Reconstructing an effective potential landscape:} From the Fokker-Planck equation, Eq.\,\ref{eq:FP_phi_2}, with natural boundary conditions, $J(\rho,\phi_2)=0$, we can obtain the steady-state solution $\rho=\pi$, satisfying $\partial_t \pi = 0$, as,

\begin{align*}
    \pi(\phi_2) = \exp \left[\int \frac{F(\phi_2)-\partial_{\phi_2}D(\phi_2)}{D(\phi_2)} d\phi_2\right].
\end{align*}
Writing the steady-state distribution as a Boltzmann factor \cite{Horsthemke2006},

\begin{align*}
    \pi(\phi_2) \propto e^{-\beta V(\phi_2)},
\end{align*}
with $\beta=1$, we can identify an effective potential landscape $V(\phi_2)$,

\begin{align*}
    V(\phi_2) = -\int \frac{F(\phi_2)-\partial_{\phi_2} D(\phi_2)}{D(\phi_2)} d\phi_2.
\end{align*}
The same approach applies to the time-dependent stochastic model, where each window has its own local steady-state, and the effective potential landscape is time dependent, 
\begin{align*}
    V(\phi_2,t) = -\int \frac{F(\phi_2,t)-\partial_{\phi_2} D(\phi_2,t)}{D(\phi_2,t)} d\phi_2.
\end{align*}

\medskip

\noindent{{\bf Stochastic model simulations of $\phi_2$:} We simulate the dynamics using an Euler scheme with the same sampling time as the data $\delta t = 1/16\,\text{s} $ or $\delta t = 1\,\text{bout}$ for the \emph{C. elegans} N2 worms and zebrafish datasets, respectively. For the non-autonomous model, we take $F_{\tau^*,h^*}^w$ and $D_{\tau^*,h^*}^w$ of the window with a center closest to the sampled time point.

\medskip

\noindent{\bf Fine-scale Markov model simulations:} As in \cite{Costa2023markovian}, we simulate symbolic sequences by sampling the next state according to the condition probability distribution $P(s_j(t+\tau^*)|s_i(t))$, which is simply the $i$-th row of $P_{ij}(\tau^*)$. From this symbolic sequence, we can then obtain a simulated time series of $\phi_2(t)$ sampled on a timescale $\tau^*$.

\medskip

\noindent{\bf Estimating the first passage time distributions:} We estimate the time spent either performing either state (``runs'' and ``pirouettes'' for \emph{C. elegans} and ``cruising'' and ``wandering'' for zebrafish) by identifying segments where $\phi_2(t)<0$ (runs/cruising) or $\phi_2(t)>0$ (pirouettes/wandering). To remove short-time fluctuations we sub-sample the data and the simulated time series by $\tau^*/2$.

\medskip

\noindent{\bf Empirical estimate of the connected auto-correlation function:} We estimate the connected autocorrelation function from $M$ time traces at each lag $\tau=l\delta t$, as

\begin{align}\label{eq:C_emp}
    \hat{C}_x(l \delta t) = \frac{1}{M}\sum_{\alpha=1}^M  \frac{\frac{1}{N-l} \sum_{i=1}^N x_{\alpha,i} x_{\alpha,i+l} -  \left( \frac{1}{N} \sum_{i=1}^N x_{\alpha,i} \right)^2}{\frac{1}{N} \sum_{i=1}^N x_{\alpha,i}^2 - \left( \frac{1}{N} \sum_{i=1}^N x_{\alpha,i} \right) ^2},
\end{align}
where $x_{\alpha,i}$ is the $i$-th frame of the $\alpha$ trace with length $N = T_\text{expt}/\delta t$. 

\medskip

\noindent{\bf Estimating the first passage time distribution of a Poisson process with varying hopping rates:} We sample $s$ according to the Boltzmann distribution $p(s)\propto \exp\left(-\frac{(s-\mu_s)^2}{2T_s}\right)$, and convert it to a hopping rate $\omega(s)$ by numerically integrating the backward Kolmogorov equation, Eq.\,\ref{eq:tau_backwards_eq}. We then sample first passage time events according to Eq.\,\ref{eq:f(t,w)}, until reaching the measurement timescale $T_\text{expt}$. We repeat this process 50,000 times, and collect the statistics of waiting times to build a normalized histogram of first passage times with logarithmic bins, which we show in Fig.\,S3.

\medskip

\noindent{\bf Estimating the first passage time distribution in the slowly-driven double well potential:} We generate 10,000 simulations of Langevin dynamics of Eq.\,\,\ref{eq:(x,s)_DW}, through an Euler-scheme with a sampling time of $\delta t = 10^{-3}\,\text{s}$ for $T_\text{expt}=10^7\,\text{s}$. We then vary $\tau_s$ in the range $[10^{-4}T_\text{expt},10^4T_\text{expt}]$ and $T_s$ in the range $[T_x/4,2T_x]$, where $T_x=10^{-3}$ and $\mu_s = \sqrt{T_x}$. The initial conditions $x(0)$ are sampled randomly either as $x(0)=1$ and $x(0)=-1$ with equal probability and $s(0) \sim \mathcal{N}(\mu_s,\sqrt{T_s})$ is sampled according to the Boltzmann distribution. From the simulations of $x(t)$, we then estimate the first passage time distribution by first identifying all the segments, $[t_0,t_f]$, in which $t_0$ corresponds to the first time $x$ returns to $x_0 = \pm 1$ after reaching $x_f=0$, and $t_f$ is the time first to reach $x_f=0$ after $t_0$. Finally, we build a normalized histogram of first passage times with logarithmic bins, which we show in Figs.\,\ref{fig:4},S4.

\medskip

\noindent{\bf Estimating the autocorrelation functions in the slowly-driven double well potential:}  We generate 50,000 simulations of Langevin dynamics of Eq.\,\ref{eq:(x,s)_DW}, through an Euler-scheme with an initial sampling time of $dt = 0.2\,\text{s}$ that is downsampled to $\delta t = 100\,\text{s}$ for $T_\text{expt}=10^8\,\text{s}$, with $T_x = 10^{-2}$, $\mu_x = \sqrt{T_x}$, $\tau_s$ sampled in the range $[10^{-2}T_\text{expt},10^2 T_\text{expt}]$ and $T_s$ sampled in the range $[1.25 T_x, 10T_x]$. We then estimate the connected autocorrelation function from the simulations using Eq.\,\ref{eq:C_emp}. The non-connected correlation function is estimated as,

\begin{align*}
    \hat{\tilde{C}}_x(l \delta t) = \frac{1}{M}\sum_{\alpha=1}^M \frac{1}{N-l} \sum_{i=1}^N x_{\alpha,i} x_{\alpha,i+l},
\end{align*}
and then normalized by dividing by $\hat{\tilde{C}}_x(l=1)$. The finite-size corrections to the correlation function are detailed in Appendix F.

\setcounter{equation}{0}

\renewcommand{\theequation}{B\arabic{equation}}

\section*{Appendix B: A primer on first passage times}

We here give a brief introduction to the concept of first passage times, following \cite{Hanggi1990,Redner2001,Pavliotis2014,Choi2020}. Broadly speaking, the first passage time probability is the probability that a random walker \emph{first} reaches a particular site at a specified time. This concept is broadly relevant in biology: from the kinetics of on and off states in membrane receptors, to the binding and unbinding of transcription factors to DNA, to the dynamics of conformation changes in proteins, to the firing of neurons and so on. 

We here focus on the particular case of escape from a metastable state, which corresponds to first passage time to reach the transition point. These are rare events: there is a timescale separation between the relaxation dynamics within a metastable state, and the time it takes to cross the transition point. The average rate of escape from a metastable state (or the mean first passage time to reach the transition point $x_f$) is given by a constant $\omega$. The distribution of the first passage times then corresponds the probability of not escaping at each infinitesimal time step until time $t$, multiplied by the probability of escaping at time $t$. The probability to not jump until time $t$ is given by $(1-\omega\delta t)^{t/\delta t}$, since at each $\delta t$ the probability of not transitioning is $1-\omega\delta t$. Taking the limit of infinitesimal time steps gives $\lim_{\delta t \rightarrow 0}(1-\omega\delta t)^{t/\delta t} = e^{-\omega t} $. Multiplying this by the probability of escaping at time $t$, the first passage time distribution in this simple case is given by $f(t,\omega) = \omega e^{-\omega t}$.

To determine $\omega$, we use the backwards Kolmogorov equation to find the mean first passage time, or mean exit time, which is defined as the expectation value of the first time $x(t)$ leaves a domain $D$, conditioned on $x(t) = x_0 \in D$. Let $\rho(y,t|x_0)$ represent the probability distribution of an ensemble of particles not leaving the domain $D$ at time $t$. It satisfies the Fokker-Planck equation with absorbing boundary conditions,

\begin{align}\label{eq:FP_fpt}
    \frac{d\rho}{dt} = \mathcal{L} \rho,\hspace{0.5cm}\rho(y,0|x_0) = \delta(y-x_0), \hspace{0.5cm} \rho|_{\partial D}  =0,
\end{align}
where $\mathcal{L}$ is the Fokker-Planck operator of the forward Kolmogorov equation. The total survival probability of still being inside the domain $D$ at time $t$ is obtained by integrating over the domain $D$,

\begin{align*}
    S(x_0,t) = \int_D \rho(y,t|x_0) dy.
\end{align*}
Given the absorbing boundary conditions at the domain boundaries, this probability is a decreasing function of time,

\begin{align*}
    \frac{\partial S}{\partial t} = - f(x_0,t),
\end{align*}
where $f(x_0,t)$ is the first passage time distribution. The mean first passage time is the first moment of this distribution,

\begin{align*}
    \overline{\tau}(x_0) & = \int_0^{\infty} f(x_0,t)\,t \, dt = \int_0^\infty S(x_0,t)dt = \\
    & = \int_0^\infty\int_D \rho(y,t|x_0) dy\,dt.
\end{align*}
Writing the solution to Eq.\,\ref{eq:FP_fpt} as $\rho(y,t|x_0) = e^{\mathcal{L} t} \delta(y-x_0)$, we get,

\begin{align*}
    \overline{\tau}(x_0) & = \int_0^\infty\int_D e^{\mathcal{L} t} \delta(y-x_0) dy \,dt \\
    & = \int_0^\infty \left( e^{\mathcal{L}^* t} 1 \right)  (x_0) dt. 
\end{align*}
Applying the adjoint operator $\mathcal{L}^*$ to the above equation we obtain,

\begin{align*}
    \mathcal{L}^*\overline{\tau}(x_0) & = \int_0^\infty \left( \mathcal{L^*}e^{\mathcal{L}^* t} 1 \right)  dt = \int_0^\infty \frac{d}{dt}\left(e^{\mathcal{L}^* t} 1 \right)  dt \\
    \mathcal{L}^*\overline{\tau}(x_0) & = -1. 
\end{align*}
For a simple concrete example, consider the dynamics of Eq.\,\ref{eq:x} with $s_t = s\in \mathbb{R}$ constant. In this case, we can use the corresponding Fokker-Planck operator (see Eqs.\,\ref{eq:FP_phi_2},\ref{eq:phi2_dot},\ref{eq:x}) to obtain, 

\begin{align*}
    \mathcal{L}^*\overline{\tau}(x_0) = -\tau_x^{-1}\partial_x U(x) \partial_x \overline{\tau}(x_0) + T_x\tau_x^{-1}\partial_x^2 \overline{\tau}(x_0) = -1 \\
    \partial_x\left(e^{-U(x)/T_x}\partial_x \overline{\tau}\right) = -\frac{e^{-U(x)/T_x}}{\tau_xT_x}.
\end{align*}
Let's consider that the particle starts at a minimum of the potential well $x_0=a$ and escapes when reaching the barrier situated at $x_f = b$. Assuming reflective boundary conditions at $x_0=a$ and an absorbing boundary at $x_f=b$ (where the escape events occurs), we have,

\begin{align*}
    \overline{\tau}(x_0) = \frac{1}{\tau_x T_x} \int_x^b e^{U(y)/T_x}dy \int_a^y  e^{-\beta U(z)/T_x} dz.
\end{align*}
Once the particle has reached $x=b$, there is a 50\% chance that it manages to escape, yielding an escape rate,

\begin{align}\label{eq:tau_backwards_eq}
    \omega^{-1} = \frac{2}{\tau_x T_x} \int_x^b  e^{U(y)/T_x} dy\int_a^y e^{-\beta U(z)/T_x} dz.
\end{align}
When the barrier height is large when compared to the level of fluctuations, $\Delta U = U(x=b)-U(x=a) \gg T_x$, we can solve these integrals asymptotically to find,

\begin{align*}
    \omega = \frac{\omega_a \omega_b}{2\pi}e^{-\Delta U / T_x} = \omega_0 e^{-\Delta U / T_x} ,
\end{align*}
where $\omega_a$ and $\omega_b$ correspond to the absolute value of second derivative of the potential at $x=a$ and $x=b$ and stem from a Taylor expansion around the minimum and the maximum of the potential, respectively. We have thus recovered Kramers approximation \cite{VanKampen1992,Hanggi1990}, Eq.\,\ref{eq:Kramers}.

\setcounter{equation}{0}

\renewcommand{\theequation}{C\arabic{equation}}

\section*{Appendix C: First passage time distribution in slowly fluctuating potential landscapes}

We here derive the expression for the first passage time distribution (FPTD) in a fluctuating potential landscape. As discussed in the main text, we consider the adiabatic limit in which the FPTD can be approximated by

\begin{align*}
    f(t) %&\sim \int_{\omega_\text{min}}^{\omega_\text{max}} p(\omega)\omega f(t,\omega) d\omega \nonumber \\
    &\propto \int_{\omega_\text{min}}^{\omega_\text{max}} p(\omega)\omega^2 e^{-\omega t}d\omega,
\end{align*}
where $\omega=\omega_0 \exp\left\{-\Delta U(s)/T_x \right\} \Rightarrow s(\omega) = \Delta U^{-1}(-T_x\log\left(\omega/\omega_0\right))$, and $\omega_0$ is a typical (fast) frequency of the hopping dynamics \footnote{We note that for the general dynamics of Eqs.\,(2,3), $\omega_0$ may have a $s$ dependency. However, without loss of generality, we consider that $\omega_0$ and $\Delta U(s)$ can be redefined to move the $s$ dependency to the exponential as a sub-dominant contribution.}. The distribution $p(\omega)$ obeys $p(\omega) d\omega = p(s) ds$, where $p(s) \propto \exp\left\{-V(s)/T_s\right\}$, and is thus given by

\begin{align*}
    p(\omega) \propto &\exp\left\{-\frac{V(s(\omega))}{T_s}\right\} \frac{T_x/\omega}{\partial_s \Delta U(s)}\,.
\end{align*}
Plugging this into Eq.\,\ref{eq:f(t)}, we get
\begin{align}\label{eq:f(t)_V_U}
    f(t) \propto &\int_{\omega_\text{min}}^{\omega_\text{max}} \exp\left\{-\frac{V(s(\omega))}{T_s}\right\}\frac{T_x}{\partial_s \Delta U(s)}\omega e^{-\omega t} d\omega.
\end{align}
The exponential factor $e^{-\omega t}$ restricts the contributions to $\omega\sim 1/t$, which motivates the change of variable $\omega = \frac{\theta}{t}$. The above integral is then recast in the form  
\begin{align}\label{eq:f(t)_all}
    f(t) \propto t^{-2} \int_{\theta_\text{min}(t)}^{\theta_\text{max}(t)} \frac{\exp\left\{-\theta-\frac{V(s(\theta))}{T_s}+\log\left(\theta\right)\right\}}{\partial_s \Delta U(s(\theta))}\,d\theta,
\end{align}
where $s(\theta)=\Delta U^{-1}\left(-T_x\log\left(\frac{\theta}{\omega_0 t}\right)\right)$, 
$\theta_\text{min}(t)=\omega_\text{min}t$ and $\theta_\text{max}(t)=\omega_\text{max}t$.

To grasp the structure of the integral, it is convenient to consider first the special case where $V$ and $\Delta U$ can be written as a power series expansion $V(s)\sim a s^n$ and $\Delta U(s) \sim b s^n$, $a,b\in\mathbb{R}$ with an equal dominant (at large values of the argument, see below) exponent $n$.  The integral reduces then to the form
\begin{align*}
    f(t) \propto t^{-2-\frac{a T_x}{b T_s}}\int_{\theta_\text{min}}^{\theta_\text{max}} \frac{\theta^{1+\frac{a T_x}{b T_s}}e^{-\theta}}{\left(-\log\left(\frac{\theta}{\omega_o t}\right)\right)^{1-\frac{1}{n}}}\,d\theta\,.
\end{align*}
It remains to verify that the time dependencies at the denominator of the integrand and the limits of integration do not spoil the behavior at large times. This is verified by noting that the numerator of the integrand has the structure of an Euler-$\Gamma$ function of order $2+\frac{a T_x}{b T_s}$. The numerator of the integrand has its maximum at $\theta^*=1+\frac{a T_x}{b T_s}$, decays over a range of values of order unity (we consider $\frac{a T_x}{b T_s}$ to be small or of order unity)  and vanishes at the origin. In that range, the argument of the power at the denominator $\log\left(\omega_0t\right)-\log\left(\theta\right)\simeq 
\log\left(\omega_0t\right)$, which yields the final scaling with subdominant logarithmic corrections
\begin{equation}\label{eq:scalingkeqn}
        f(t) \sim t^{-2-\frac{a T_x}{b T_s}} \times \log(\omega_0 t)^{\frac{1}{n}-1}.
\end{equation}
To complete the argument, we note that the time dependency of $\theta_\text{min}$ is not an issue as long as values $\theta\sim O(1)$ are in the integration range. In practice, this means that the minimum hopping rate $\omega_\text{min}$ should be comparable to (or larger than) the measurement time, $\omega_\text{min}^{-1} \sim O(T_\text{expt})$.

Before moving to the general case, two remarks are in order. First, for $\omega_0 t\gg 1$ the functions $V$ and $\partial_s \Delta U$ that appear in Eq.~\ref{eq:f(t)_all} have their argument $s\gg 1$. The dominant behavior of the two functions should then be understood for large values of their arguments. Second, the denominator $\partial_s \Delta U$ could {\it a priori} be included in the exponential at the numerator but this does not modify our conclusion. It is indeed easy to verify that the maximum $\theta^*$ and the decay range would not be shifted at the dominant order (and this holds also for the general case considered hereafter).

We can now consider the general case with different dominant exponents $V(s)\sim a s^n$ and $\Delta U(s) \sim b s^k$, $a,b\in\mathbb{R}$. The argument of the exponential in Eq.~\ref{eq:f(t)_all}
\begin{equation}
\label{eq:Lagr}
L\left(\theta\right) = -\theta-\frac{V(s(\theta))}{T_s}+\log\left(\theta\right)\,,
\end{equation}
has its maximum at $\theta^*$, defined by the implicit equation
\begin{align*}
\theta^* = 1+\frac{T_x}{T_s}\frac{\partial_s V(s\left(\theta^*\right))}{\partial_s \Delta U(s\left(\theta^*\right))}=1+\frac{T_x}{T_s}\frac{an}{bk}s^{n-k}\,,
\end{align*}
where we have used 
\begin{align*}
    \partial_\theta V(s) =
    \partial_s V(s)\times\frac{ds(\theta)}{d\theta}\,;\quad \frac{ds(\theta)}{d\theta} = -\frac{T_x/\theta}{\partial_s \Delta U(s)}\,.
\end{align*}

For $n<k$, the maximum $\theta^*\simeq 1$ (as $s\gg 1$) and the integrand decays in a range of order unity. Indeed, the dominant order of the derivatives $\partial^p L$ ($p\ge 2$) at $\theta=\theta^*$ coincide with those of $\log\left(\theta\right)$. It follows that $L(\theta)-L(\theta^*)\simeq \log\left(\theta/\theta^*\right)-\left(\theta-\theta^*\right)$. The resulting integral over $\theta$ is an Euler $\Gamma$-function of order two, which indeed forms at values $O(1)$. In that range, $s\sim \left(\frac{T_x}{b} \log\left(\omega_0 t\right)\right)^{1/k}$ and the integral is then approximated by $\exp\{{L}\left(\theta^*\right)\}$, so $f(t)$ becomes
\begin{align*}
    f(t) \sim t^{-2}\exp\left\{-\frac{a\left(\frac{T_x}{b}\log(\omega_0 t)\right)^{n/k}}{T_s}\right\} \,.
\end{align*}
The factor at the denominator in Eq.~\ref{eq:f(t)_all} is $O\left[\exp\left\{\left(1/k-1\right)\log[\log(\omega_0 t)]\right\}\right]$ and thus of the same order as terms that we have discarded in our approximation so we neglect it as well. Since the integral over $\theta$ forms for values $O(1)$, the constraint on the minimum hopping rate is the same as for the $n=k$ case, i.e., $\omega_\text{min}^{-1} \sim O(T_\text{expt})$.

For $n>k$, the maximum $\theta^*\sim \left(\log\omega_0 t\right)^{n/k-1}$, which is now large. The dominant order of the derivatives $\partial^p{  L}$ ($p\ge 2$) at $\theta=\theta^*$ is given by $\left(-1\right)^{p-1}\left(p-1\right)!\left(\theta^*\right)^{-(p-1)}$, that is they coincide with those of $\theta^*\log\left(\theta\right)$. It follows that
${  L}(\theta)-{  L}(\theta^*)\simeq \theta^*\left[ \log\left(\theta/\theta^*\right)-\left(\theta-\theta^*\right)/\theta^*\right]$. The resulting integral over $\theta$ is an Euler $\Gamma$-function of (large) argument $\theta^*+1$\,: its value is approximated by Stirling formula, which yields $\int \left(\theta/\theta^*\right)^{\theta^*}e^{-\left(\theta-\theta^*\right)}\,d\theta\simeq \sqrt{\theta^*}$. The $\sqrt{\theta^*}$ reflects the fact that the integral forms around the maximum at $\theta^*$ of the integrand over a range $\sqrt{\theta^*}$, which implies that the approximation $-\log\left(\frac{\theta}{\omega_0t}\right)\simeq \log\left(\omega_o t\right)$ still holds, as in the previous cases $n\le k$.
The $\sqrt{\theta^*}$, as well as the $\log\left(\omega_0 t\right)^{1/k-1}$ coming from the denominator in Eq.~\ref{eq:f(t)_all}, are subdominant with respect to terms that we have neglected in the expansion of ${  L}$. We therefore discard them from our final approximation for $n>k$\,: 
\begin{align*}
    f(t) \sim t^{-2}\exp\left\{-\frac{a\left(\frac{T_x}{b}\log(\omega_0 t)\right)^{n/k}}{T_s}\right\}\,.
\end{align*}
Since the integral over $\theta$ forms for values $O\left(\left(\log\omega_0 t\right)^{n/k-1}\right)\gg 1$, the condition $\omega_\text{min}^{-1} \sim O(T_\text{expt})$ ensures {\it a fortiori} that the finite value of $\omega_\text{min}$ does not affect the above result.

Discarding subdominant terms, in all three cases we thus get the general expression we present in the main text,
\begin{equation}\label{eq:f(t)_(k,n)}
    f(t) \sim t^{-2}\exp\left\{-\frac{a\left(\frac{T_x}{b}\log(\omega_0 t)\right)^{n/k}}{T_s}\right\}\,.
\end{equation}

To verify the validity of the above arguments, we show in Fig.\,S6 how, to the dominant order, asymptotic predictions agree with a detailed numerical integration of Eq.\,\ref{eq:f(t)_V_U} for $\Delta U(s)=s^k$ and $V(s) = s^n$. 

\setcounter{equation}{0}

\renewcommand{\theequation}{D\arabic{equation}}

\section*{Appendix D: First passage time for a Poisson process with varying hopping rates}

To probe our theoretical predictions, we first assume that hopping events are well captured by a Poisson process and that the modulation of the potential landscape is infinitely slow such that the adiabatic approximation of Eq.\,\ref{eq:f(t)} holds exactly. In practice, we sample $s$ according to the Boltzmann distribution, Eq.\,\ref{eq:Boltmzann_s}, and, in order to relax from the Kramers approximation, we obtain $\omega(s)$ directly through numerically integrating the backward Kolmogorov equation, Eq.\,\ref{eq:tau_backwards_eq} (with $\tau_x=1$) \cite{VanKampen1992}.

We then sample events according to the distribution of first passage times $f(t,\omega)$, Eq.\,\ref{eq:f(t,w)}, until reaching the measurement time $T_\text{expt}$ (see Fig.\,S3(a) for a schematic of the sampling procedure). We take $U(x,s) = s^2(x^2-1)^2$ to be a symmetric double well potential and sample $s$ according to a Boltzmann distribution with $p(s) \propto \exp\left\{ - \frac{(s-\mu_s)^2}{2T_s}\right\}$, corresponding to an Ornstein-Uhlenbeck process. Since the first passage time distribution is dominated by the large energy barriers, $s \gg \mu_s$, we take $V(\Delta U^{-1}(x)) \sim x/2$. From the derivation of Eq.\,\ref{eq:f(t)_asymp}, we expect that the final distribution of first passage times will be given by,

\begin{equation}
   f(t) \sim t^{-2-\frac{T_x}{2T_s}}\,.
\end{equation}
Indeed, this is what we find through numerical simulations, Fig.\,S3(b).

\setcounter{equation}{0}

\renewcommand{\theequation}{E\arabic{equation}}

\section*{Appendix E: Correlation functions in slowly fluctuating potential landscapes}

We here derive the expression for the correlation function in a fluctuating potential landscape. In general, the autocorrelation function can be expressed as a sum over exponential functions \cite{Risken1989,Coffey2004},

\begin{align*}
    C_x(\tau) = \frac{\left\langle x(t)x(t+\tau)\right\rangle - \langle x \rangle ^2}{\langle x^2 \rangle - \langle x \rangle^2}  = \sum_i c_i e^{-\Lambda_i \tau}
\end{align*}
where $\Lambda_i$ ($\Lambda_1<\Lambda_2<\ldots$) are the eigenvalues of the Fokker-Planck operator and $\sum_i c_i=1$. Since we are interested in the long-term behavior of systems with energy barriers that can fluctuate over time, we assume that the large $\tau$ behavior of the correlation function asymptotes to

\begin{align*}
    C_x(\tau) \sim e^{-\Lambda_1 \tau}\,,
\end{align*}
where $\Lambda_1$ is the first non-trivial eigenvalue, which captures the longest-lived dynamics in the system. In addition, we assume that there is always a deeper well, with an escape rate $\omega$, that dominates the long-lived dynamics. In this case, $\Lambda_1 \propto \omega$ and we have,

\begin{align*}
    C_x(\tau) \sim e^{-\omega \tau}\,.
\end{align*}
As previously discussed, we take the adiabatic approximation to derive the asymptotic behavior of the correlation function in the presence of slow non-ergodic modulation of the potential landscape. In particular, we obtain a weighted average of the correlation function over multiple realizations of $\omega(s)$, yielding

\begin{align}
    C_x(\tau) \sim \int_{\omega_\text{min}}^{\omega_\text{max}} p(\omega) e^{-\omega \tau}d\omega.
\end{align}
Note that in comparison with Eq.\,\ref{eq:f(t)}, besides dropping an $\omega$ factor due to the difference between $f(t,\omega)$, Eq.\,\ref{eq:f(t,w)}, and $C_x(\tau)$, we also do not need to take into account the extra factor of $\omega$ coming from the finite observation time, which in the case of the estimation of first passage times biases the probability density in a manner that is proportional to $\omega$. In the case of the correlation function, the dynamics of $x$ is exposed to modulations in $s$, regardless of $\omega(s)$. Following the same steps as before, we consider that $V$ and $\Delta U$ can be written as a series expansion with dominant terms $V(s)\sim a s^n$ and $\Delta U(s) \sim b s^k$, $a,b\in \mathbb{R}$. In this case, we find that to dominant order

\begin{align}
    C_x(\tau) \sim \exp\left\{-\frac{a\left(\frac{T_x}{b}\log(\omega_0 \tau)\right)^{n/k}}{T_s}\right\}.
\end{align}
Notably, when $n=k$, we obtain power law correlations with an exponent that depends on the ratio of temperatures,
\begin{align}\label{eq:Cx_asymp}
    C_x(\tau) \sim \tau^{-\frac{a T_x}{b T_s}} \times \log(\omega_0 \tau)^{\frac{1}{n}-1},
\end{align}
where we have included sub-dominant corrections. In particular, we find that when $n=k$ both the first passage time distribution $f(t)\sim t^{\beta}$ and the correlation function  $C_x(\tau) \sim \tau^\gamma$ have power-law behavior at large times, and that the exponents are related by $\gamma = \beta + 2$. In addition, when $T_s\rightarrow \infty$ correlations decay slowly as $C_x(\tau) \sim \log(\omega_0 \tau)^{1/n-1}$.

\setcounter{equation}{0}

\renewcommand{\theequation}{F\arabic{equation}}

\section*{Appendix F: Finite-size correction to the correlation function}

When estimating the correlation function from a collection of $M$ finite time traces sampled at $\delta t$ and with length $N = T_\text{expt}/\delta t$, we compute,

\begin{widetext}
    \begin{align*}
    \hat{C}_x(\tau = l \delta t) = \frac{1}{M}\sum_{\alpha=1}^M  \frac{\frac{1}{N-l} \sum_{i=1}^N x_{\alpha,i} x_{\alpha,i+l} -  \left( \frac{1}{N} \sum_{i=1}^N x_{\alpha,i} \right)^2}{\frac{1}{N} \sum_{i=1}^N x_{\alpha,i}^2 - \left( \frac{1}{N} \sum_{i=1}^N x_{\alpha,i} \right) ^2},
\end{align*}
\end{widetext}

where $x_{\alpha,i}$ is the $i$-th frame of the $\alpha$ trace.  Assuming that the finite-size corrections to the correlation function are dominated by corrections to the mean value (and not the variance), we can leverage the derivation of Desponds et al. \cite{Desponds2016} to obtain an expression for the finite-size corrections to the correlation function from the non-connected correlation function $\tilde{C}_x = \langle x(t) x(t+\tau) \rangle$,

\begin{widetext}
    \begin{align}\label{eq:C_correction}
    C_c(\tau) & \sim  \tilde{C}(l)+\frac{1}{N}\left(\frac{1}{N}-\frac{2}{N-l}\right)\left(N\tilde{C}(0) + \sum_{k=1}^{N-1}2(N-k)\tilde{C}(k) \right) + \nonumber \\
    & \frac{2}{N(N-l)}\left(l\tilde{C}(0)+\sum_{k=1}^{l-1}2(l-k)\tilde{C}(k)+\sum_{m=1}^{N-1}\tilde{C}(m)(\min(m+l,N)-\max(l,m)) \right).
\end{align}
\end{widetext}

As detailed in the main text, as a case study we take the overdamped dynamics for the position $x$ of a particle in a symmetric double well potential, for which the barrier height can fluctuate according to a slow parameter $s$, Eq.\,\ref{eq:(x,s)_DW}. The time scale separation between the hopping events and the relaxation to the well means that the correlation function is dominated by the first non-trivial eigenvalue, $C_x(\tau) \sim e^{-\Lambda_1 \tau} = e^{-2\omega \tau}$, where we take $\Lambda = 2\omega$ due to the fact that the potential wells have the same depth \cite{Risken1989}. Taking $V(s)\sim s^2/2$ and  $\Delta U(s) = s^2$ we obtain that, in the asymptotic large $\tau$ limit, $C_x(\tau) \sim \tau^{-\frac{T_x}{2T_s}} \log(\omega_0 \tau)^{-1/2}$, Fig.\,5(a).

To obtain an accurate estimate of the correlation function for all $\tau$, we go beyond the asymptotic approximation and numerically integrate

\begin{align}\label{eq:C_tilde_DW}
    \tilde{C}_x(\tau) \sim \int_{-\infty}^{\infty} e^{-V(s)/T_s} e^{-2\omega(s)\tau} ds,
\end{align}
where $\omega(s)$ can be estimated directly by integrating the Kolmogorov backward equation. At large $\tau$, the numerical integration of $\tilde{C}_x(\tau)$ matches the asymptotic behavior $\tau^{-\frac{T_x}{2T_s}}$. Plugging Eq.\,\ref{eq:C_tilde_DW} into Eq.\,\ref{eq:C_correction}, we obtain the correction to the autocorrelation function $C_c(\tau)$ presented in Figs.\,5(b,c). 

\bibliography{bibliography}

\clearpage

\onecolumngrid

\setcounter{figure}{0}
\setcounter{page}{1}
\setcounter{equation}{0}

\makeatletter
\renewcommand{\theequation}{S\arabic{equation}}
\renewcommand{\thefigure}{S\arabic{figure}}
\renewcommand{\thepage}{S\arabic{page}}

\section*{Supplemental Material}

\begin{figure*}[ht!]
    \centering
    \includegraphics{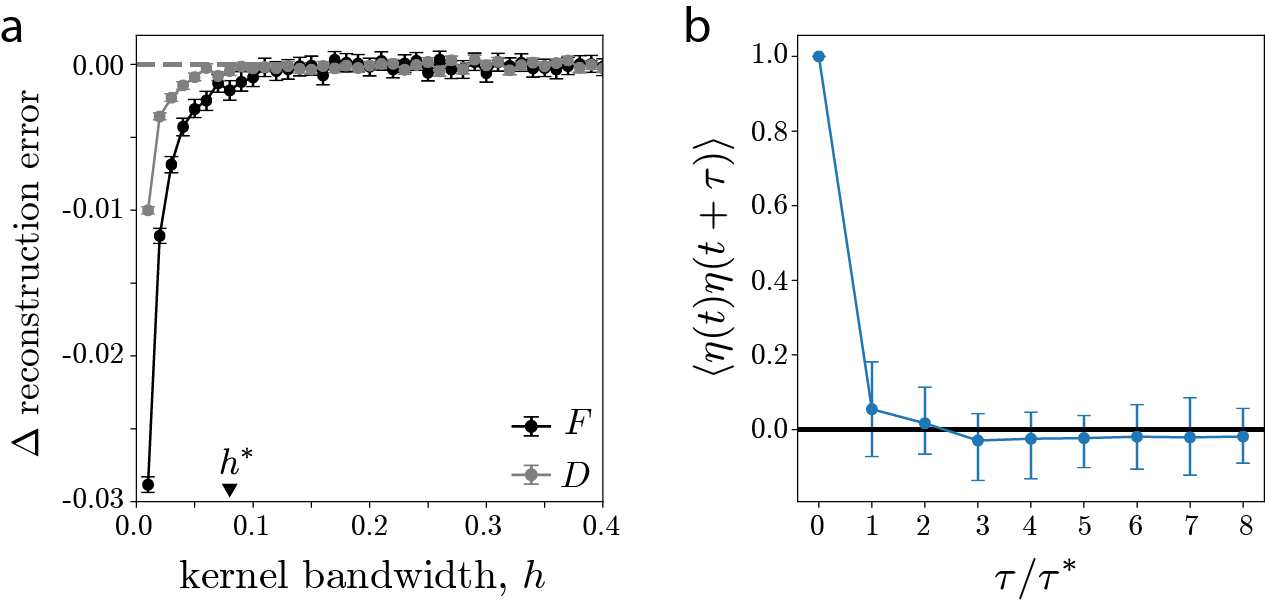}
    \caption{{\bf Details of the stochastic model inference in \emph{C. elegans} wild type worms.}
    (a) Change in reconstruction error $\Delta \xi = \xi(h+\delta h) - \xi(h)$  of the drift $F$ and diffusion $D$ coefficients depending on the bandwidth of the kernel used to perform the Kramers-Moyal averages \cite{Lamouroux2009} (see Appendix A). For each value of bandwidth $h$, we simulate trajectories using the estimated drift and diffusion coefficients. We then re-infer the drift and diffusion coefficients from the simulated trajectories and compare them against the parameters obtained from the original time series to get the reconstruction error, Eq.\,\ref{eq:rec_error} ($\Delta$-algorithm in \cite{Lamouroux2009}).  We chose $h^* = 0.08$ as the lowest $h$ value when the reconstruction error stops changing (when $\Delta\,\text{reconstruction error} \approx 0$). 
    (b) Autocorrelation function of the residuals $\eta(t)$ after fitting Eq.\,\ref{eq:phi2_dot} to time series of $\phi_2(t)$. At the sampling time $\tau^*$ the noise decorrelates, thus justifying the white noise approximation. Error bars correspond to 95\% confidence intervals bootstrapped over 1000 simulations of randomly sampled worms.
    }
    \label{fig:S_inference}
\end{figure*}

\begin{figure*}[ht!]
    \centering
    \includegraphics{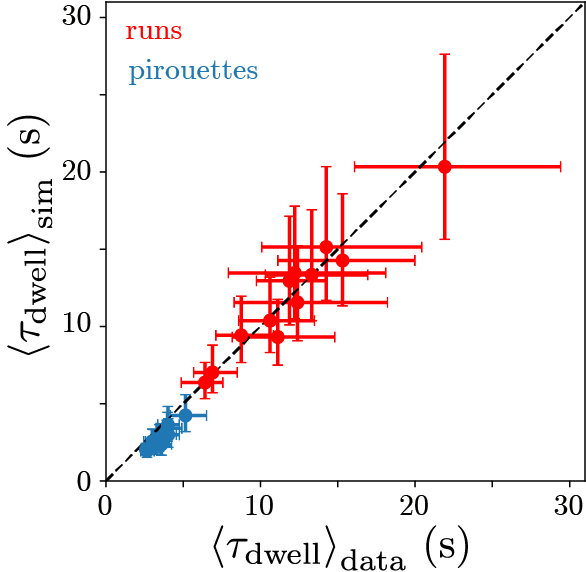}
    \caption{{\bf Simulations from Eq.\,\ref{eq:phi2_dot} accurately predict the average time spent in a given behavioral state for \emph{C. elegans} wild type worms}. Each point corresponds to a worm in a particular state, and the error bars correspond to 95\% confidence intervals bootstrapped over events.
    }
    \label{fig:S_static_model}
\end{figure*}

\begin{figure*}
   \centering
   \includegraphics{ 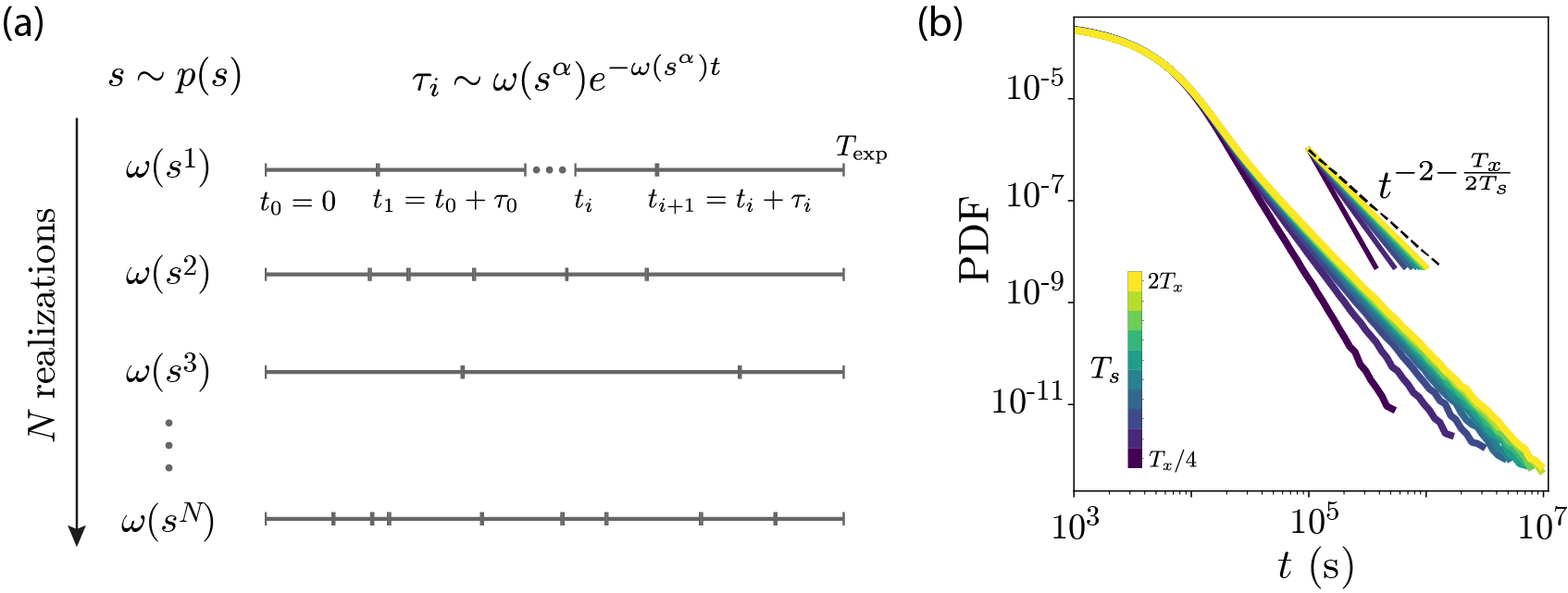}
   \caption{{\bf Heavy-tailed first passage time distribution in Poisson process with varying hopping rates.}
   (a) Schematic of the simulation process. For each realization, we sample $s$ according to the Boltzmann distribution, $p(s)$. The hopping rate corresponding to a particular sample $s^i$ is then determined by the backward Kolmogorov equation, Eq.\,\ref{eq:tau_backwards_eq}, and event durations are sampled according to the first passage time distribution $f(t,\omega) = \omega e^{-\omega t}$ until reaching the experimental timescale $T_\text{expt}$. This process is then repeated over $N=$50,000 realizations (see Appendix A).
   (b) Probability density function (PDF) of first passage times for the Poisson process with varying hopping rates. As predicted, we obtain a power law with an exponent $f(t)\sim t^{-2-\frac{T_x}{2T_s}}$.
   }
   \label{fig:S_poisson}
\end{figure*}

\begin{figure*}[ht!]
    \centering
    \includegraphics{ 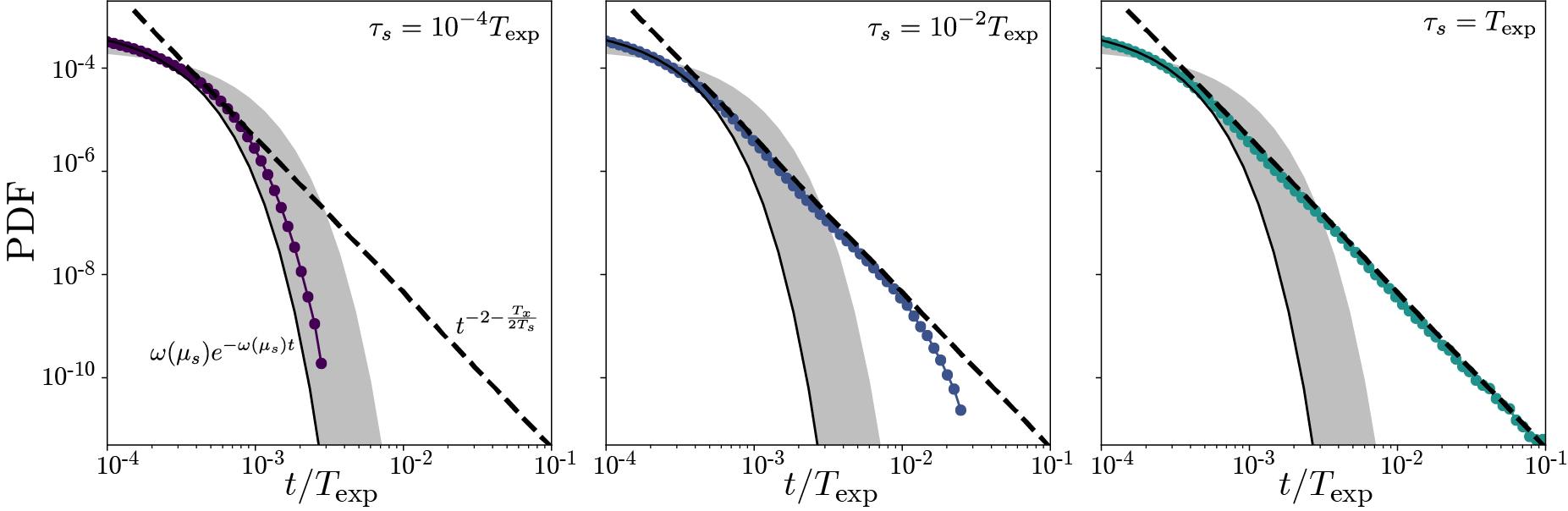}
    \caption{{\bf Emergence of power-law tails in the first passage time distributions for the slowly-driven double well dynamics of Eq.\,\ref{eq:(x,s)_DW} as a function of $\tau_s$.}
    For $\tau_s\ll T_\text{expt}$ (left) the potential relaxes to its mean value faster than the hopping timescale, resulting in exponential behavior with a decay rate corresponding to the mean value $\omega(\mu_s)$. The black dashed line and gray shaded area correspond to $f(t,\omega)=\omega e^{-\omega t}$ with $\omega = \omega(\mu_s)$ and $\omega(\mu_s + \sqrt{T_s})$ respectively. As $\tau_s$ increases, the regime in which we observe power law behavior grows, and for intermediate $\tau_s$ we obtain a truncated power law with an exponential tail starting at $t\sim\tau_s$ (middle). Finally, when $\tau_s = T_\text{expt}$ the measured tail of the distribution is power-law distributed (right). The black dashed line corresponds to our prediction $t^{-2-\frac{T_x}{2T_s}}$ and $T_s = T_x/2$. 
    }
    \label{fig:S_DW}
\end{figure*}

\begin{figure*}[ht!]
    \centering
    \includegraphics{ 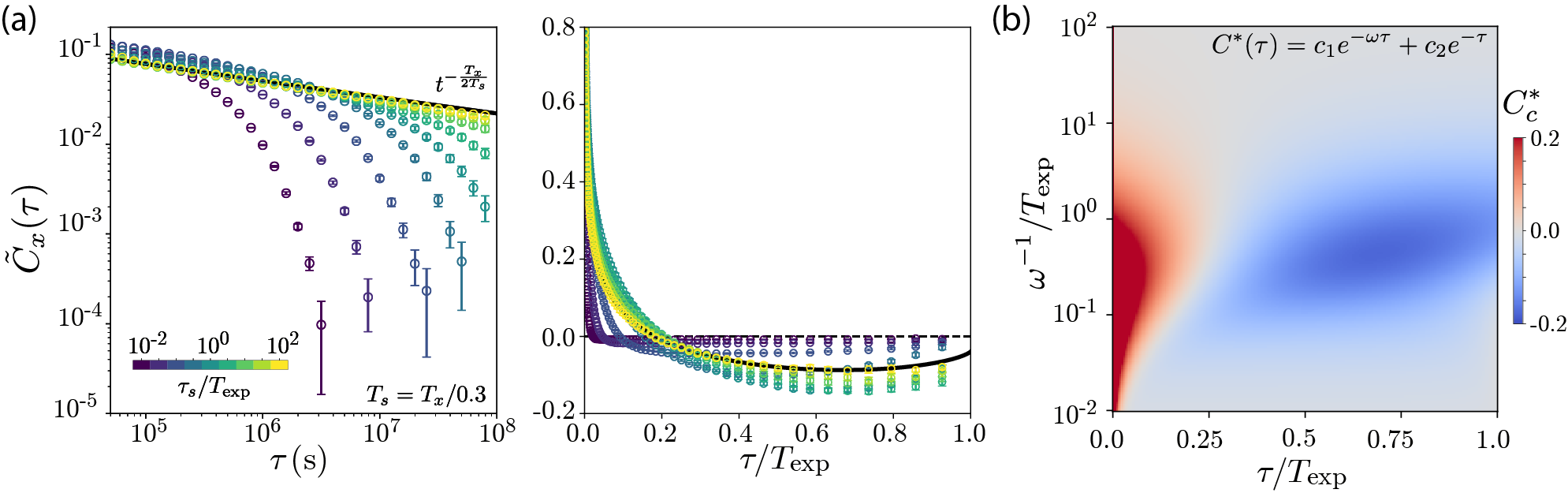}
    \caption{{\bf Dependence of the autocorrelation function on $\tau_s$.}
    (a-left) Non-connected correlation function for $T_s = T_x/0.3$ and varying $\tau_s$. When $\tau_s$ is small, the tail of the correlation function is exponential, with a decay rate that is given by the maximum barrier height attained at this temperature. As $\tau_s$ grows, the variation of the potential landscape becomes slower and slower and our asymptotic approximation correctly predicts the emergent power law behavior at large $\tau_s$.
    (a-right) Connected correlation function for $T_s = T_x/0.3$ and varying $\tau_s$, estimated directly from the time series data. Our adiabatic approximation correctly predicts the correction to the correlation function for large $\tau_s$. We normalize both correlation functions by their value at $\tau = 1\,\text{lag} = 5 \times 10^{-4} T_\text{expt}$. Notably, the correlation function exhibits finite-size corrections even when it has exponential tails, as long as the timescale of the exponential decay is comparable to the measurement time scale. Error bars represent 95\% confidence intervals bootstrapped across 50,000 simulations.
    (b) We examine the finite-size correction to the correlation function when there is no time dependence to the hopping rate $\omega$. For a system with a single fixed energy barrier, we would expect that the correlation function would be given by $C_x(\tau) = c_1 e^{-\omega  \tau} + c_2 e^{-\tau}$ \cite{Risken1989,Coffey2004}, where $\sum_i c_i = 1$ and we take the intrinsic time scale of relaxation to a well to unity without loss of generality. As expected, even when the correlation function has exponential tails we observe the appearance of finite-size effects when $0\ll \omega^{-1} \lesssim T_\text{expt}$. Notably, when $\omega^{-1}\rightarrow \infty$ these finite-size effects are less apparent since only the short timescale survives. In contrast, when we allow the hopping rate to fluctuate in time we effectively generate a continuum of time scales such that, even when $T_s\rightarrow\infty$, finite-size effects are still apparent, Fig.\,\ref{fig:5}(c).
    }
    \label{fig:S_DW_acfs}
\end{figure*}

\begin{figure*}[ht!]
    \centering
    \includegraphics{ 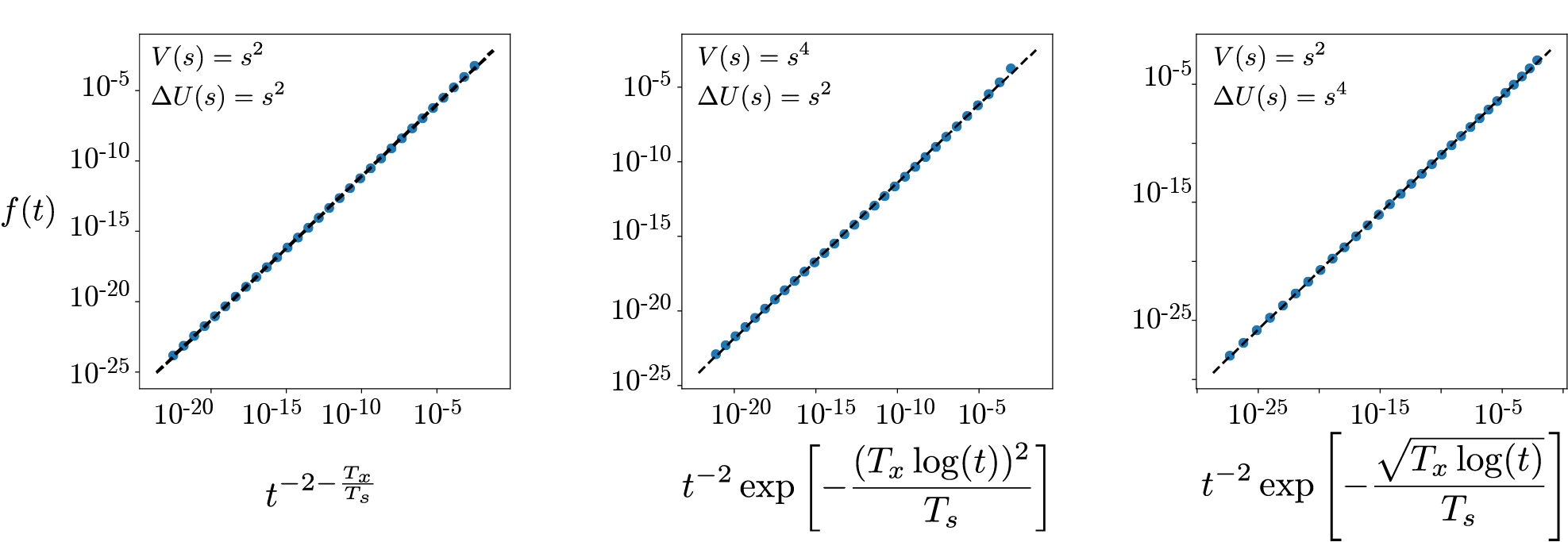}
    \caption{Numerical integration of $f(t)$ for different choices of $V(s)$ and $\Delta U(s)$, compared to the asymptotic approximation of Eq.\,\ref{eq:f(t)_(k,n)} (black dashed line) with $T_x = 0.1$ and $T_s = 0.2$. We numerically integrate Eq.\,\ref{eq:f(t)_V_U} with $\Delta U(s)=s^k$ and $V(s) = s^n$, through a Riemann sum using the midpoint rule from $\omega_\text{min} = 5\times 10^{-10}$ to $\omega_\text{max}=1$ with $\Delta \omega = 10^{-9}$.
    }
    \label{fig:asymptotics}
\end{figure*}

\begin{figure*}
    \centering
    \includegraphics{ 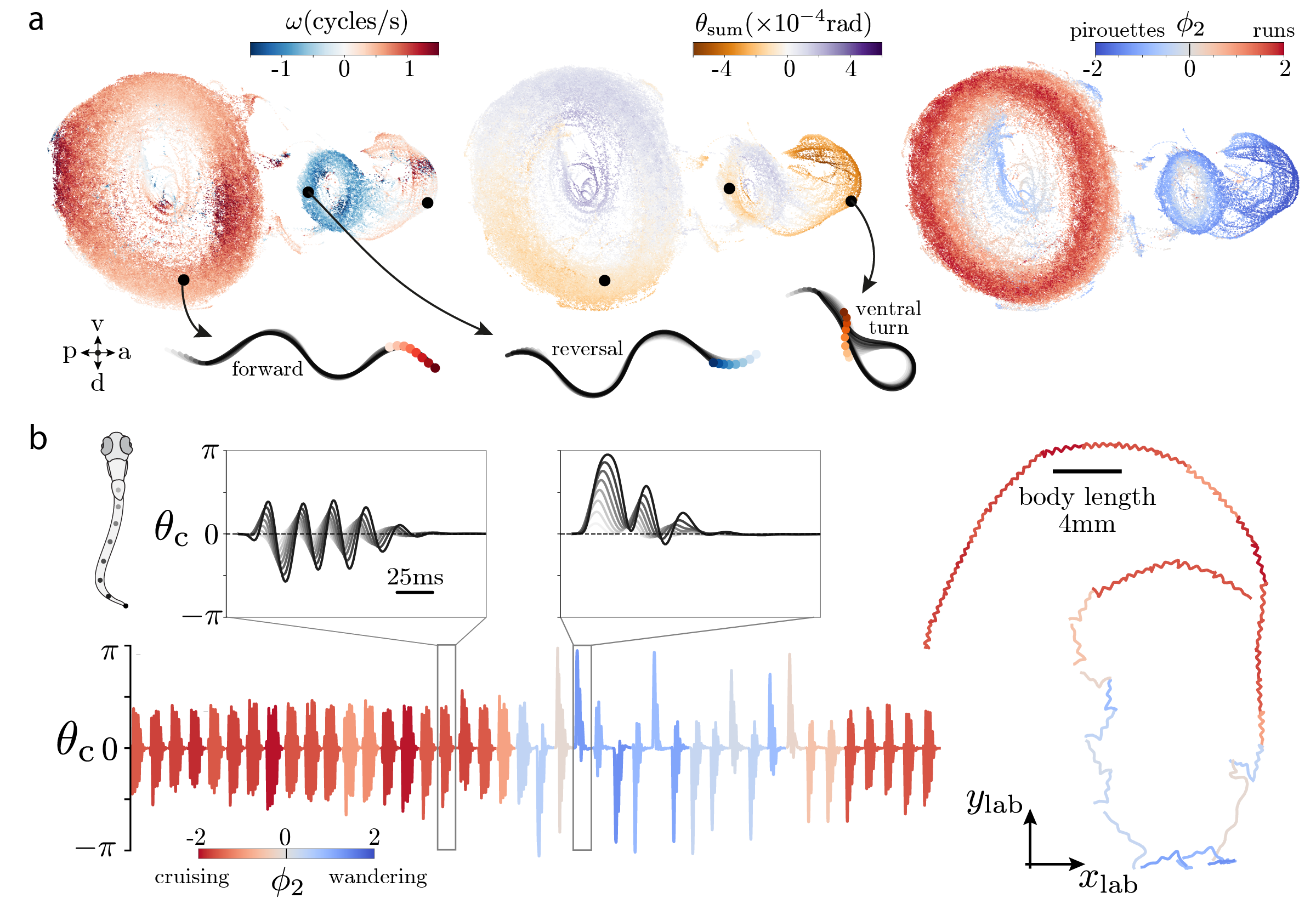}
    \caption{ {\bf Details of the analysis of posture dynamics in \emph{C. elegans} \emph{npr-1} mutants and larval zebrafish.}
    (a) From videos of \emph{npr-1} mutants crawling on a food-rich environment, we proceed as in Fig.\,\ref{fig:2}, extracting the body posture (using {\tt wormpose} \cite{Hebert2021}) and reconstructing a maximally-predictive space of posture sequences \cite{Costa2023markovian,Costa2023} with $K^*=0.4\,\text{s}$ (see Appendix A). We show a 2-dimensional projection of the state-space, obtained through UMAP, and color code this space either according to the body wave phase velocity $\omega$ \cite{Stephens2008} (left), the overall body curvature (sum of the tangent angles along the body) $\theta_\text{sum}$ (middle), and the projection along the slowest eigenvector of the inferred transition matrix $\phi_2$ (right), obtained with $\tau^* = 0.5\,\text{s}$ (see Appendix A). In addition, we show a few example sequences of postures $X_{K^*}$, illustrative of different behaviors: forward, reversal and ventral turn (colored circle indicates the head-position, and light-to-dark color indicates the passage of time). Notably, we observe that the overall dynamics is somewhat similar to what is observed in wild type N2 worms off-food (see Fig.1 in \cite{Costa2023markovian}), except for the notable absence of dorsal turns. As in Fig.\,\ref{fig:2}, we find that $\phi_2$ captures transitions between forward ``runs'' and combinations of reversals (negative $\omega$) and turns (large $\theta_\text{sum}$) that constitute ``pirouettes''. We obtain $K^*$, $N^*$ and $\tau^*$ in the same way as for the wild type dataset (see Appendix A).
    (b) We collected data from \cite{Groneberg2020} in which larval zebrafish are exposed to a chasing dot stimulus for $5\,\text{s}$ every $2\,\text{min}$ for at least 1 hour. The fish move in discrete tails bursts (bouts), interrupted by periods in which the tail is immobile. A custom tracking algorithm used in \cite{marques2018structure} identifies each bout and collects the position of 9 points along the tail at $700\,\text{Hz}$ during each bout. The posture is represented as the cumulative tail angle $\theta_c$ from head to tail, sampled for $175\,\text{frames}$ (enough to capture the tail's relaxation). From the bout dynamics, we proceed as for \emph{C. elegans} \cite{Costa2023markovian}: we identify maximally-predictive sequences of $K^*=5$ bouts, and infer the slow dynamics $\phi_2$ though the eigenspectrum of the inferred Markov chain with $\tau^*=3\,\text{bouts}$ (see Appendix A). We plot an example sequence of $\approx 40\,\text{bouts}$, in which the cumulative tail angles are color coded by the projection along $\phi_2$, as well as the resulting trajectory of the head position (right). The slow dynamics $\phi_2$ corresponds to a ``wandering-cruising'' axis, in which the fish either engages in bout sequences with large orientation changes (``wandering''), or performs sequences of smoother forward bouts (``cruising'') that result in more persistent trajectories.
    }
    \label{fig:S_npr-1_zebrafish}
\end{figure*}

\end{document}